**Title:** Simulated phase dependent spectra of terrestrial aquaplanets in M-dwarf systems

**Authors:** E.T. Wolf[1,2,4], R.K. Kopparapu[2,3,4,5], J. Haqq-Misra[2,5]

## Affiliations:

[1]University of Colorado, Boulder, Laboratory for Atmospheric and Space Physics, Department of Atmospheric and Oceanic Sciences, Boulder CO, USA

[2]Virtual Planetary Laboratory, Seattle WA, USA

[3]NASA Goddard Space Flight Center, Greenbelt MD, USA

[4]Sellers Exoplanet Environments Collaboration, NASA Goddard Space Flight Center, Greenbelt MD, USA

[5]Blue Marble Space Institute of Science, Seattle WA, USA

**Abstract:**

Orbital phase dependent variations in thermal emission and reflected stellar energy spectra can provide meaningful constraints on the climate states of terrestrial extrasolar planets orbiting M-dwarf stars. Spatial distributions of water vapor, clouds, and surface ice are controlled by climate. In turn water, in each of its thermodynamic phases, imposes significant modulations to thermal and reflected planetary spectra. Here, we explore these characteristic spectral signals, based on 3D climate simulations of Earth-sized aquaplanets orbiting M-dwarf stars near the habitable zone. By using 3D models, we can self-consistently predict surface temperatures and the location of water vapor, clouds, and surface ice in the climate system. Habitable zone planets in M-dwarf systems are expected to be in synchronous rotation with their host star and thus present distinct differences in emitted and reflected energy fluxes depending on the observed hemisphere. Here, we illustrate that icy, temperate, and incipient runaway greenhouse climate states exhibit phase dependent spectral signals that enable their characterization.

## 1. Introduction

Planets in M-dwarf star systems provide our best chance for characterizing habitable terrestrial exoplanetary atmospheres in the next decade. Observing planets orbiting M-dwarf stars has several key positive biases favoring observation. M-dwarf stars are small and dim. Their low stellar luminosities and small radii mean that planetary emitted and reflected radiances will be relatively easier for humans to observe, compared to those from an Earth-like planet orbiting a Sun-like star. Habitable zone planets around M-dwarfs must also orbit quite close to their host stars, and thus their orbital periods will be much less than an Earth-year, allowing for observations to be feasibly stacked over many orbits within nominal mission lifetimes.

Habitable extrasolar planets are assumed to have significant amounts of liquid water on their surfaces and thus in contact with the atmosphere. This provides a quantifiable constraint on the temperature regimes and climate states allowable for habitable worlds. Nominally, the surface temperature of a habitable planet must be maintained between the freezing and boiling points of water, however, locally surface temperatures may violate these limits. Broadly surface habitable worlds generally must maintain global mean surface temperatures somewhere between 280 K and 310 K, or else they risk having widespread areas of severe temperature extremes (Wolf et al. 2017).

The presence of water vapor, perhaps uniquely among atmospheric constituents, acts as an inherently destabilizing force in the climate system. In its vapor phase, water-vapor is both a strong greenhouse gas, absorbing thermal emission by the planet, and also a significant absorber of stellar radiation at near infrared wavelengths. Both of these features of water vapor spectroscopy, combined with the temperature dependence of the Clausius-Clapeyron relationship, accelerate temperature perturbations (*i.e.* a positive climate feedback). The thermal absorption component is commonly referred to as the water-vapor greenhouse feedback. The near-infrared stellar absorption component also represents a positive feedback, and is greatly accentuated for planets around M-dwarf stars, given their low effective temperatures and thus peak emission in the near-infrared region of stellar spectra (Kopparapu et al. 2013). Furthermore, water also readily condenses at typical atmospheric temperatures and pressures for habitable worlds, forming clouds, sea-ice, and snow, all of which can add significantly to the planetary albedo. In particular, the temperature dependence of sea-ice and snow surface coverage also accelerates temperature perturbations, through the sea-ice albedo feedback. However, the spectral reflectivity of water ice and snow decreases significantly into the near-infrared which mutes the sea-ice albedo feedback for planets around M-dwarf stars (Joshi et al. 2011; Shields et al. 2013; Shields et al. 2014). Clouds have competing effects depending on their location. Liquid water clouds that form in the boundary layer (*i.e.* stratus clouds) contribute greatly to the planetary albedo, however ice water clouds that form in the upper troposphere (*i.e.* cirrus clouds) contribute to the greenhouse effect while remaining optically thin to incoming stellar radiation.

Due to their close proximity to the host star, habitable zone planets around late-K and M-dwarf stars are expected to be tidally locked into synchronous rotation states (Kasting et al. 1993). For a synchronously rotating planet, its rotation rate equals its orbital period, and the same side of the planet always faces the host star. Numerous climate models have explored these novel atmospheric states for terrestrial extrasolar planets (e.g. Joshi et al. 1997; Merlis & Schneider 2010; Yang et al. 2013; 2014; Hu & Yang 2014; Shields et al. 2016,

2018; Kopparapu et al 2016, 2017; Fujii et al. 2017; Turbet et al. 2017, 2018; Haqq-Misra et al. 2018; Del Genio et al. 2019; Komacek & Abbot 2019; Yang et al. 2019; Adams et al. 2019). The Coriolis force plays a central role in modulating the atmospheric circulation, and the assumption of synchronous rotation allows precise constraints of planetary rotation rates. For slowly rotating planets ($P \lesssim 20$ Earth days), the severe weakening of the Coriolis force allows for a thermally direct circulation, with strong rising motions on the permanent day-side and sinking air on the permanent night-side. Such a circulation tends to produce thick reflective clouds on the substellar hemisphere, greatly increasing the planetary albedo, and stabilizing the climate against a runaway greenhouse under relatively high total stellar irradiances (Yang et al. 2013). However, the longitudinal extent of the substellar clouds is sensitive to the specific rotation rate. As the planetary rotation rate increases for habitable zone planets around late M-dwarfs, strengthening Coriolis forces drive increased zonal motions that advect clouds eastward around the planet (Kopprapu et al. 2017; Haqq-Misra et al. 2018).

Water vapor, clouds, and ice are variable constituents in the climate system of habitable planets, while other typical gas species (*i.e.* $N_2$, $CO_2$) tend to be well-mixed. Here, we postulate that the spatial variability of water in its various thermodynamic phases, coupled with its strong interaction with thermal and stellar radiation, provides an avenue for atmospheric characterization through phase dependent observations. Broadband phase curves and phase dependent spectra may provide an opportunity to characterize atmospheric circulation, heat transport, and cloud formation in planetary atmospheres, beyond what may be capable with transit and thermal emission spectroscopy alone (Koll & Abbot 2016). Phase dependent information captures multi-dimensional process in planetary atmospheres, which are particularly relevant for tidally locked planets around M-dwarf stars, where day-night asymmetries in climate and atmospheric structure are much more likely to occur (e.g. Joshi 2003; Yang et al. 2013; Koll & Abbot 2016; Haqq-Misra et al. 2018). Along with increasing the albedo, the presence of thick substellar clouds on slowly rotating planets can also significantly reduce the outgoing longwave radiation from the substellar hemisphere due to the greenhouse effect of high altitude clouds (Yang et al. 2013). For habitable planets, thermal emission phase curve morphology is controlled by emission from the planet surface and its interaction with water-ice clouds in the atmosphere, which are controlled by atmospheric circulations in the upper troposphere/lower stratosphere region (Haqq-Misra et al. 2018). This stands in contrast to the case of hot Jupiters, where thermal emission phase curve morphology is thought to be controlled by direct stellar heating of the atmosphere and subsequent advection of heat eastward by broad equatorial superrotating jets (e.g. Knutson et al. 2007; Showman & Polvani 2011).

Here, we analyze the phase dependence of thermal emission and reflected stellar energy from Earth-sized ocean covered planets. We examine both broadband fluxes and spectral radiances. Our focus is in how phase dependent planetary energy signals are modulated by water vapor, liquid water clouds, ice water clouds, and surface ice. Planetary energy fluxes and spectra are determined based on a suite of 3D theoretical models of Earth-sized aquaplanets located in and near the habitable zones of late-K and M-dwarf stars. We use the habitable zone planet models of Kopparapu et al. (2017), along with complimentary simulations that extend to lower stellar fluxes and thus to colder planetary temperatures.

This paper marks the third in a series of papers on 3D climate modeling of terrestrial planets around M-dwarf stars. Kopparapu et al. (2017) describes the climatology and habitability constraints of planets approaching the moist and runaway greenhouse limits. Haqq-Misra et al. (2018) describes the atmospheric circulation patterns as a function of rotation rate and orbital distance. This work focuses on thermal and reflected phase curves from these worlds, and how phase curve information may be used to characterize terrestrial exoplanetary atmospheres in M-dwarf systems. This manuscript is organized as follows. In section 2 we outline our methodology. In section 3 we briefly summarize climate modeling results. In section 4 we describe broadband thermal emission and broadband planetary albedo phase curves. In section 5 we describe phase dependent thermal emission and reflection spectra. In section 6 we describe the variation spectra. Sections 7 and 8 contain discussions and conclusions respectively.

## 2. Methods

We follow the methods used in the general circulation modeling study of the inner edge of the habitable zone for synchronous rotators by Kopparapu et al. (2017) except where noted. Here we give a brief summary of our modeling methods, but refer the reader to section 2 in Kopparapu et al. (2017) and Neale et al. (2010) for progressively in-depth model descriptions. We use a modified version of the Community Atmosphere Model (CAM) version 4 from the National Center for Atmospheric Research in Boulder, CO (Neale et al. 2010). Specific code changes, including radiative transfer, model configurations, and initial conditions files needed to facilitate exoplanet studies with CAM, are available via the `ExoCAM` and `ExoRT` code packages, which we have made publicly available at the lead authors' Github page[1]. In this work, planets are assumed to be Earth-sized, completely ocean covered, and with atmospheres composed of 1 bar $N_2$ and variable amounts of water vapor, liquid water clouds, and ice water clouds. There is no $CO_2$, $CH_4$ or other non-condensable (at Earth-like temperatures) greenhouse species in these simulations. Water vapor and clouds are prognostically determined using the standard parameterizations of CAM version 4 (Neale et al. 2010; see also Kopparapu et al. 2017, section 2.3). For simplicity and computational efficiency, the ocean is treated as a 50-meter deep "slab" ocean with no ocean heat transport (*i.e*. q-fluxes are set to zero everywhere). Sea ice forms where sea-surface temperatures fall below the freezing point of sea-water at 271.36 K. We assume the default snow and ice albedo parameterization, which divides ocean, sea ice, and snow albedos into two bands: visible and near-IR divided at ~0.7 μm. Following Shields et al. (2013), we set the visible (near-IR) sea-ice albedo to 0.67 (0.3), the snow albedo to 0.8 (0.68), and the ocean albedo to 0.07 (0.06), respectively. This two-band parameterization of surface albedos is hard-wired within sea-ice, snow, and land modules respectively, and is not easily changed to accommodate multiple bands. A band resolved surface albedo parameterization would surely improve its treatment (e.g. Joshi et al. 2003), however this is left for future work. Simulations were run with $4° \times 5°$ horizontal resolution and 40 vertical layers extending from the surface up to ~1 mb pressures, using a finite volume dynamical core (Lin & Rood, 1996).

[1] https://github.com/storyofthewolf

The radiative transfer module considers $H_2O$ line and continuum absorption and $N_2$-$N_2$ collision-induced absorption (CIA). Correlated-K absorption coefficients are derived for $H_2O$ using the HITRAN 2012 database (Rothman et al. 2013) and `HELIOS-K`, an ultrafast open-source spectral sorting program that runs on graphics processing units (Grimm & Heng 2015). Water vapor continuum is treated using the formalism of Paynter & Ramaswamy (2011). $N_2$-$N_2$ CIA is included from Borysow & Frommhold (1986). Cloud overlap is treated using the Monte Carlo Independent Column Approximation assuming maximum random overlap (Pincus et al 2003). In this study, the radiative transfer module is configured with 68 spectral intervals across the whole spectra, each with 8 gauss points per spectral interval. Note, here we have increased the spectral resolution compared to Kopparapu et al. (2017), which used only 42 spectral intervals. Differences in the radiative flux calculations between the 42 and 68 bin versions are less than ~1% at worst for moist greenhouse atmospheres, and introduce no discernable differences into the mean states of the climate simulations. Here, we have chosen to use the higher resolution radiative transfer in order to extract higher quality synthetic spectra from our simulated worlds. Longwave and shortwave radiative streams share the same spectral interval grid, and were computed each over the entire available spectrum. We have modified our 3D model to directly output spectrally resolved fluxes.

We simulated planets around stars with stellar effective temperatures, $T_{eff}$, of 2600 K, 3000 K, 3300 K, 3700 K, 4000 K, and 4500 K, assuming BT-SETTL spectra (Allard et al. 2007), with stellar metallicities of [Fe/H] = 0.0, and log $g$ = 4.5 (Figure 1). The planets are assumed to be in synchronous rotation. The incident stellar fluxes and orbital periods (and thus planetary rotation rates) are calculated with consistency to Kepler's laws. For stars with $T_{eff}$ ≥ 3300 K, we use the empirically derived stellar mass-luminosity relationships from Boyajian et al. (2013). For cooler stars we use stellar mass-luminosity relationships derived from stellar evolutionary models (Baraffe et al. 2002), assuming an age of ~3 Gyr. We have duplicated the simulation grid of Kopparapu et al. (2017), and have also included two new simulations per star, with 680 Wm$^{-2}$ (0.5 $S_0$) and 1080 Wm$^{-2}$ (0.8 $S_0$) incident stellar fluxes respectively. We included these new simulations to the grid in order to expand our sampling of theoretical climate states to cold worlds. In total, around each stellar type, we have conducted simulations of terrestrial planets with global mean surface temperatures ($T_s$) varying from ~205 K up through the point where a runaway greenhouse is triggered.

Based on our suite of 3D model simulations, we then compute shortwave and longwave phase curves and phase dependent spectra directly from 3D GCM spectrally resolved outputs, using the methods described by Koll and Abbot (2015, Appendix C) for calculating the disk-averaged and observer-projected fluxes from a planet, assuming an edge-on view relative to the observer. Because our 3D model directly outputs spectrally resolved information, no off-line radiative calculations are needed. For climatologically stable simulations, phase curves and phase dependent spectra are calculated based on averages of the last 5 Earth-years of simulation. Our testing has found that a 5 year duration for a temporal average is sufficient for removing bias from variability. For incipient runaway greenhouse cases, as described in Kopparapu et al. (2017, section 3.3), temperatures rapidly increase and the model becomes numerically unstable when model temperatures approach ~400 K within ~10 Earth-years time in most cases. For incipient runaway greenhouse cases, we calculate phase curves based on the last 90 Earth-days of simulation before the numerical instability occurs. This short duration for a temporal average is chosen as to sample the hottest part of the simulation, but as a consequence retains some variability due to weather. Here, we limit the scope of our focus to planets that are also transiting, thus variations in the observed phase curves caused by observer obliquity are not considered here. Planets that are transiting will provide the best targets, allowing us to combine information between thermal and reflected phase curves, transit spectroscopy, secondary eclipse mapping, and radial velocity methods.

## 3. Summary of Climate Models

Most of the climate models used here were first presented in Kopparapu et al. (2017), with continued analysis in Haqq-Misra et al. (2018). Here we have expanded our simulation grid to include lower stellar fluxes (0.5 $S_0$ and 0.8 $S_0$), in order to include cold and icy states in our ensemble of climates. We refer the reader to Kopparapu et al. (2017) for an in-depth discussion of planetary climate, clouds, and moist and runaway greenhouse processes occurring near the inner edge of the habitable zone. We refer the reader to Haqq-Misra et al. (2018) for an in-depth discussion of atmospheric dynamical regimes that exist on these tidally locked worlds as a function of changing planetary rotation rate. In this section we summarize the climate modeling results that are most relevant for interpreting phase dependent thermal emission and reflected stellar energy spectra from these worlds.

We first divide climate into four different categories, sorted by their global mean surface temperature, $T_s$. These categorizations will be referenced throughout the paper. Cold climates have 200 K < $T_s$ < 250 K. While these worlds are very cold and their surfaces are dominated by sea ice, all still maintain a small fraction (4% – 25%) of open-ocean at the substellar point and thus are habitable based on the standard criteria requiring surface liquid water to exist anywhere on the planet surface. Cool climates have 250 K < $T_s$ < 273 K. These climate states have open-ocean fractions up to ~50%, thus their substellar hemispheres have little to no ice, while their antistellar hemispheres are permanently ice-covered. Temperate climates have 273 K < $T_s$ < 310 K and represent the most optimistic cases for habitable planets, dominated by open-oceans and largely with modern Earth-like surface temperatures. Finally, incipient runway states are those where a runaway greenhouse instability has been triggered. As described in Kopparapu et al. (2017, section 3.3), a strong energy imbalance is present and model temperatures rapidly increase, however our climate model becomes numerically unstable as temperatures near ~400 K. Simulating a fully realized runaway greenhouse, with atmospheric temperatures beyond the critical point of water, is beyond the operational bounds of our code and remains a significant challenge for 3D climate modelers (e.g. Ishiwatari et al. 2002; Kane et al 2018). A fully realized runaway greenhouse would evaporate the entirety of the oceans into its atmospheres, yielding ~270 bars of water vapor for every Earth ocean originally present (Kasting 1988), and would have a surface temperature of ≥1600 K in equilibrium (Goldblatt et al. 2013). Thus, our "runaway" cases represent a snapshot of atmospheres undergoing the runaway greenhouse process and are in a transient state, not a steady-state. Our presented runaway states fall well short of the extremes of the classical runaway greenhouse (Goldblatt & Watson 2012). With respect to observable phenomenon, our incipient runaway states are perhaps best considered as a sampling of hot, moist, and optically thick atmospheres (e.g. Goldblatt et al. 2015).

To further sort our results, we divide atmospheric circulation regimes into three categories. For the remainder of the paper we will adopt the circulation regime nomenclature suggested by Haqq-Misra et al. (2018). "Slow" rotators have rotation periods ($P$) of >20 Earth-days, which occurs for habitable zone planets around stars with 3300 K ≤ $T_{eff}$ ≤ 4500 K. Slow rotators are characterized by a weak Coriolis force, strong rising motions at the subtellar point, cooling and descending air on the antistellar hemisphere, and radially symmetric day to night side heat transport. "Rapid" rotators have $P < 5$ Earth-days, which can occur only for habitable zone planets around the coolest of stars, $T_{eff} < 2600$ K. Rapid rotators are characterized by a strong Coriolis force, weak upwelling, strong superrotation, and west to east zonal heat transport. Lastly, an intermediate dynamical regime, coined "Rhines" rotators by Haqq-Misra et al. (2018), features rotation rates of $5 \leq P \leq 20$ Earth-days, which occurs for habitable zone planets around stars with 3000 K $\leq T_{eff} \leq$ 3300 K. Defined in terms of the Rhines length scale (Rhines, 1975; Showman et al. 2010), Rhines rotators are characterized by a relatively strong Coriolis force, strong upwelling, and moderate superrotation. While here we define discrete categorizations based on specific rotation periods, in actuality atmospheric circulation evolves continuously as a function of planet rotation. Note, also it has been suggested that bistability in atmospheric circulation regimes could occur for terrestrial sized planets with rotation rates of less than ~10 days depending upon assumed boundary conditions (Carone et al. 2018), the possibilities for which we have not yet explored.

Here, we show maps of surface temperature (K, Figure 2), vertically integrated water vapor column (kg m$^{-2}$, Figure 3), vertically integrated liquid cloud water path (g m$^{-2}$, Figure 4), and vertically integrated ice cloud water path (g m$^{-2}$, Figure 5). Figures 2, 3, 4, and 5 have the same format. Each column shows illustrative cases for a cold planet (~200 K), a cool planet (~260 K), a temperate planet (~300 K), and a hot, moist planet where a transition to a runaway greenhouse has been triggered (~380 K). Each row corresponds to a different host star, with stellar effective temperatures ranging from 2600 K to 4500 K. The stellar effective temperature, range of orbital periods, and range of incident stellar fluxes are listed in the left-hand margin. Naturally, cold planets are furthest from their host star and have the longest orbital periods and lowest received stellar fluxes. Planets where a runaway greenhouse has been triggered are closest to their host star, and have the shortest orbital periods, and highest received stellar fluxes. Recall, that all of the worlds simulated are assumed to be in synchronous rotation, thus their orbital period equals the planetary rotational period.

To first order, surface temperature maps vary as expected, with climate warming as the stellar flux is increased (Figure 2). Note that day-to-night heat transport becomes more efficient as the climate warms, and thus day-night temperature gradients are reduced for increasingly warm states. For incipient runaway greenhouse states, the atmosphere is optically thick and the surface temperature varies little between substellar and antistellar hemispheres for all rotation rates. For cold climates a sharp day-night temperature gradient is maintained as the substellar point retains surface liquid water, while the night-side remains frigid. Slow rotating planets ($T_{eff}$ ≥ 3300 K, P ≥ 20 days) generally maintain surface temperature distributions that are radially symmetric about the substellar point. However for planets with faster rotation rates, found for habitable zone planets around low mass stars, increased zonal transport results in the emergence of longitudinally banded patterns.

The water vapor column also varies as expected with warming climate (Figure 3), with the amount of water vapor that the atmosphere can contain increasing exponentially as a function of temperature, as determined by the Clausius-Clapeyron relation. Recall, here we have assumed completely ocean-covered planets, which thus provide an unlimited source of water into the atmosphere. Cold planets have only a maximum of < 10 kg m$^{-2}$ of water vapor in their atmospheres centered over a small substellar patch of open-ocean, and virtually no water vapor on their ice-covered night sides. On the other hand, our incipient runaway greenhouse cases have 5000 to 10,000 kg $m^{-2}$ in their atmospheres at the point of model termination. Cool climates may have up to ~75 kg $m^{-2}$ and temperate climates have up to several hundred kg $m^{-2}$ of water vapor in their atmospheres. Note water vapor distributions are generally symmetric about the substellar point for slow rotators, but begin to take on zonal patterns due to eastward advection that occurs for faster rotators.

In Figures 4 and 5 we show liquid water cloud and ice water cloud paths respectively. Note that clouds are shown in white. While surface temperatures and the water vapor column behave predictably and monotonically with increasing incident stellar flux, the behavior of clouds is more complex. Liquid water clouds are sparse for cold climates, because these atmospheres generally remain cold and dry. However, cold climates have the highest concentrations of ice water clouds. However, unlike for Earth, where ice clouds (i.e. cirrus) are only formed high in the atmosphere and thus contribute to the greenhouse effect, here for cold climate ice clouds exist near the surface and thus do not imply a significant greenhouse effect. Liquid water clouds are thickest for cool and temperate climates. For slow rotating planets ($T_{eff}$ ≥ 3300 K) thick clouds decks form around the substellar hemisphere (e.g. Yang et al. 2013). However, note that the for planets around low mass stars, the increased planetary rotation rates for habitable zone planets results in a stronger Coriolis force, which drives zonal flows (Haqq-Misra et al. 2018) and correspondingly advects clouds eastward from the substellar hemisphere (Kopparapu et al. 2017). Incipient runaway greenhouse planets tend to have fewer clouds on the substellar hemisphere. Strong stellar irradiance and strong absorption in the near-infrared by water vapor heats the atmosphere significantly, lowering relative humidities, and resulting in fewer clouds in the substellar hemisphere despite overwhelming water vapor burdens (Kopparapu et al. 2017). Liquid water clouds are generally confined to the marginally cooler night-side, and ice-clouds are few. Maps of liquid and ice cloud water paths are critical for the interpretation of thermal emitted and reflected stellar phase curves, as will be made evident in the following sections.

# 4. Broadband Phase Curves

## 4.1. Thermal Emission

First, we compute broadband thermal emission phase curves based on all simulations conducted (Figure 6). Each panel shows top-of-atmosphere thermal emission phase curves ($Wm^{-2}$) integrated across the entire spectrum, for all simulations conducted around each star respectively. Stellar effective temperatures, $T_{eff}$, are labeled in the upper left-hand corners of each panel. Note the illustration of phases at the top of the figure. Thermal emission phase curves are color coded according to climate state, with relevant global mean temperature ranges also given. The solid lines indicate stable climate states which have reached equilibrium. Dark blue lines indicate cold climates (200 K < $T_s$ < 250 K). Light blue lines indicate cool climates (250 K < $T_s$ < 273 K). Solid gold lines represent temperate climates (273 K < $T_s$ < 310 K), our ideal habitable scenarios. The dashed red lines are for incipient runaways, where a transition to a runaway greenhouse has been triggered, and provide a representation of a hot and moist atmosphere (see Section 3).

Planets in slow, Rhines, and rapid rotating circulation regimes exhibit fundamentally different climate state dependent morphologies of their broadband thermal emission phase curves. Slow and Rhines rotators (i.e. $T_{eff} \geq 3000$ K) exhibit an inversion of their thermal phase curve amplitudes as the climate warms from cold to temperate states. As noted first by Yang et al. (2013), temperate synchronously rotating planets tend to have efficient day-night heat redistribution (Figure 2), coupled with thick clouds on the substellar hemisphere that extend high into the troposphere (Figures 4, 5). The strong greenhouse effect provided by these high substellar ice clouds creates a minimum in the observed thermal emission from the planet observed at secondary eclipse (*i.e.* viewing the day-hemisphere). The night-side of the planet remains cloud free and thus can emit thermal radiation to space originating from much lower and thus warmer layers of the atmosphere. Conversely, for cold planets, day-night temperature gradients are large with the night side remaining icy and very cold, but with a small patch of open-ocean confined to the substellar point (Figure 2). On cold planets, cloud coverage is sparse and confined to the boundary layer, where their emitting temperature remains near to that of the surface (Figure 4, 5). The result is that for cold planets the thermal emission phase curve tends to have a maximum at secondary eclipse corresponding with when the small patch of open-ocean at the substellar point comes into view. Cool planets, represent an in between case, and an additional mode in their broadband thermal phase curves is evident. On these worlds, clouds are only concentrated within ~30° of the substellar point, resulting in a local minimum at the substellar point, and emission maxima observed near western and eastern quadrature respectively.

The inversion of the thermal phase curve with warming climate is predicted for planets around stars with $T_{eff} \geq 3000$ K, including both slow and Rhines rotators. However, this trend is not predicted for rapidly rotating planets that are found in the habitable zone of 2600 K stars. Around such cool stars, all climate states yield similar thermal phase curve morphologies, with maximum emission observed near secondary eclipse. The thermal emission minimum is generally found at eastern quadrature. For rapidly rotating planets, the Coriolis effect becomes important and increases the strength of the zonal circulation. Here, superrotating winds advect clouds eastward of the substellar point, limiting their reflectivity at the substellar point and confining ice clouds to reside near the eastern terminator (Figure 5).

An additional mode of variation in thermal emission phase curves involves the phase offset of maximum and minimum emission. Slow rotators tend to maintain strict day-night symmetry of thermal emission phase curves for cold, cool, and temperate climate states, holding true even as the phase curve inverts for warm climates as described above. Thermal emission phase curves for cold climates remain symmetric for all circulation regimes, because clouds are few and emission is dominantly from the surface, whose energy budget is defined simply by the solar insolation and surface albedo. However, Rhines and fast rotators exhibit symmetry breaking of their thermal emission phase curves for warmer climates. As climates warm, clouds take a leading order importance in the determination of thermal emission. For rapid rotators, clouds, and in particular the location of high ice clouds (Haqq-Misra et al. 2018), are shifted significantly eastward of the substellar point, causing the thermal emission minimum to be observed nearer to eastern quadrature. Interestingly, for Rhines rotators, the high ice clouds are shifted westward, causing the thermal emission minimum to be observed nearer to western quadrature (Figure 5).

Lastly, incipient runaway simulations represent an outlier case, which do not conform with trends deduced from stable climate states. We find, similar to other recent studies (e.g. Leconte et al. 2013; Kane et al. 2019), that as a runaway greenhouse is triggered the atmospheric relative humidity and cloudiness decline, despite large specific humidities. The low relative humidity ensures that the magnitude of thermal emission to space generally exceeds the theoretical radiation limit from saturated atmospheres of 282 $Wm^{-2}$, determined by Goldblatt et al. (2013) from line-by-line calculations. The large water vapor burden results in strong absorption of stellar near-infrared radiation (Figure 3). The day side becomes preferentially heated in the middle atmosphere, and then can effectively reradiate thermal energy back to space. Clouds in our incipient runaway cases tend to be few and are greater on the night-side of the planet (Figure 4, 5).

Note that Earth presently has a mean outgoing longwave radiation of 239.7±3.3 $Wm^{-2}$ (Stephens et al. 2012). On Earth, the global mean outgoing longwave radiation varies by only ~4 $Wm^{-2}$ (~1%) over monthly timescales due to cloud variability (*i.e.* weather) while interannual variability is even smaller (Harris and Belotti, 2010). There would be little inferred phase dependence of thermal emission from an Earth-twin planet orbiting a Solar-twin star. Cowan et al. (2012), using 3D model simulations of Earth from the ECHAM climate model, predict little more than a 2% variation in deduced thermal phase variations for an Earth-twin planet observed at a viewing obliquity of 0°. By contrast, the phase dependent variations in thermal emission is much larger for synchronously rotating habitable planets around M-dwarfs. Here, while global mean outgoing longwave radiation remains on the same order as Earth (~200 to 300 $Wm^{-2}$), for synchronously rotating worlds the amplitude of variation can be up to ~70 $Wm^{-2}$, or roughly 20% – 35% of the global mean.

### 4.2. Albedo

In Figure 7, we show broadband albedo phase curves based on all simulations conducted. Figure 7 is arranged and color coded identically as Figure 6. Similar to broadband thermal phase curves, planets in different circulation regimes, and in different climate states, exhibit different morphologies of phase dependent albedo variation. Note, unlike thermal emission which emanates from all parts of the planet, albedo is only calculated from the illuminated fraction of the planet. At secondary eclipse the entire substellar hemisphere is illuminated relative to the observer. At crescent phases, only a small sliver of the planet near the terminator is illuminated relative to the observer. By definition the albedo at transit is undefined.

Perhaps surprisingly, for slow rotating planets ($T_{eff} \geq 3300$ K), we find that temperate and cool climates exhibit larger planetary albedos than do cold icy climates states, at all orbital phases except at crescent phases. Cold climate states generally have few clouds and are covered by sea ice, except for small regions confined to the substellar point. For cool and temperate climate states, the illuminated hemisphere has little or no surface ice, and has significant cloudiness enshrouding much of the substellar hemisphere (Figure 2, 4). Irradiated by the red spectra of M-dwarf stars (Figure 1), liquid water clouds have a greater broadband reflectivity than does sea-ice and snow (Shields et al. 2013). Thus, cloudy climates of habitable planets around M-dwarf stars will display higher albedos than will a water-ice covered "Snowball" world. For slow rotators, cold climates exhibit a strong increase in albedo at crescent phases. For cold climates, small areas of open-ocean along with cloud free conditions around the substellar point effectively suppress the disk-averaged albedos observed at secondary eclipse. However, at crescent phases, only ice-covered regions of the planet are visible to the observer, and thus exhibit a higher apparent albedo at these phase angles, compared to at secondary eclipse. For cool and temperate worlds around slow rotators, phase dependent albedos remain fairly uniform, except at western crescent phases. This may be attributed to a subtle decreases in cloudiness along western terminator, at trailing cusp of the substellar cloud deck relative to the direction of planetary rotation (Figure 4).

As we move to smaller and cooler host stars, where Rhines and rapid rotation circulation regimes are present, we note changes to albedo phase curve morphologies. As planet rotation rate increases, eastward zonal flow strengthens and water clouds are advected to the east of the substellar point (Kopparapu et al. 2017; Haqq-Misra et al. 2018). This causes a tilting of the broadband albedo phase curve, with slightly higher albedos being observed at eastern quadrature and crescent phases for habitable planets around 3000 K, and 3300 K and cooler stars. Rapidly rotating planets found around 2600 K stars again are an outlier compared to the other cases. Opposite of the behavior seen for slower rotating planets, for rapidly rotating planets the albedo monotonically increases for progressively colder planets. This is because superrotation is strong and effectively obliterates the subtellar cloud deck. Clouds are quickly sheared downstream, away from the subtellar point (Figure 4). Thus for fast rotators, the planetary albedo at secondary eclipse is controlled primarily by the surface albedo, rather than by clouds as is the case for slower rotating circulation regimes. For temperate planets around 2600 K stars, the strong advection of clouds eastward results in highest albedos found at eastern quadrature and crescent phases. Increased albedos are also found at western crescent phase, due to strong weather systems which circle the planet, carrying moisture and clouds originating from the substellar hemisphere (Figure 3, 4).

Around all stars, planets undergoing a runaway greenhouse process exhibit exceedingly low albedos. This is due to an obvious lack of sea-ice, a relative lack of clouds, and large amounts of water vapor in the atmosphere which readily absorbs stellar energy in the near-infrared. For stars with $T_{eff} \geq 3000$ K, ensuing runaway greenhouse climate states may be easily discerned via their low albedos, particularly when viewed at secondary eclipse. Clouds are generally isolated to the terminators and night-side of the planet, where the free atmosphere is marginally cooler than the highly irradiated substellar hemisphere, allowing water vapor to condense into cloud (Figure 4, 5). Increases in the phase dependent albedo at crescent phases can be contributed to these clouds located near the terminators. For fast rotators around 2600 K stars, the differentiation between runaway and temperate climates is less obvious, as albedos in both cases remain low and of similar magnitude. Still phase dependent changes to the albedo may yield a tell-tale sign. For temperate planets, tropospheric water clouds are present but are advected eastward by strong supperotating winds, resulting in the highest reflection at eastern quadrature. However, for incipient runaway cases, the advection of heat eastward maintains high temperatures, low relative humidities, and few clouds in the eastern hemisphere. Clouds preferentially form in the subtly cooler western hemisphere, resulting in the highest reflection being observed at western crescent phases (Figure 4). Similar patterns of cloud formation in hot, supperrotating atmospheres, with more cloud mass found in the cooler western hemisphere, have also been noted in models of hot Jupiters (Parmentier et al. 2016). Finally, note that for all cases considered, there is a general trend that for a given climate state the planetary albedos are smaller around cooler stars, because the incident stellar light becomes increasingly shifted into the near-infrared (Figure 1), where both water vapor and sea ice are significant absorbers.

## 5. Phase Dependent Spectra

### 5.1. Thermal Emission Spectra

In this section we shift our focus to spectrally resolved phase curves. In Figure 8 we show thermal emission spectra measured at the top-of-atmosphere, from cold, temperate, and ensuing runaway greenhouse climates states (columns), around each of our 6 stars (rows). Spectra are shown for the 4 primary phases; transit (dark blue), western quadrature (gold), secondary eclipse (red), and eastern quadrature (light blue).

In all cases, we see negligible phase dependent variation in thermal emission within water vapor absorption bands. Phase dependent differences are small at long wavelengths (<15 μm) where water vapor rotational bands imply a relatively smooth absorption spectra, and they are also small coincident with the~6 μm water vapor absorption band. At even shorter wavelengths (<5 μm), the magnitude of phase dependent thermal emission differences is damped due to there being very little planetary emission. The most significant phase dependent spectral variations in thermal emission are found in the water vapor window region, which extends approximately between 8 – 13 μm. The water vapor window region is also coincident with maximum in thermal emission from terrestrial planets at roughly Earth-like temperatures. In the water vapor window region, the spectral signals of clouds and their variability are manifested most strongly, because there are no other absorbers in this spectral region. The presence and location of water clouds is driven to first order by the planet temperature and atmospheric circulation regime, and thus can yield clues as to the planet's climate state. Recall in our model, the only absorbers are water vapor, clouds, and $N_2$-$N_2$ CIA. $N_2$-$N_2$ CIA only weakly affects thermal radiation longward of ~20 μm and is likely swamped by $H_2O$ absorption anyway (Wordsworth & Pierrehumbert, 2013).

Note, $N_2$-$N_2$ CIA also has an absorption feature at 4.3 μm which may be detectable at <10 part-per-million levels in transmission spectra (Schwieterman et al. 2015), however it lies at the shoulder of both stellar and planetary energy distributions and thus has minimal effects on reflection and emission spectra, and does not affect climate.

Cold planets yield phase dependent thermal emission spectra that are quite similar, regardless of the host star and atmospheric circulation regime. For cold planets, there is very little water vapor and few clouds in the atmosphere, and thus the effects of constituent advection are less important. Thermal emission in the water vapor window region is dominated by emission from near the surface directly to space. Clouds that are present in cold cases are confined to close to the surface, and thus do not impose a significant greenhouse effect. The spatial distribution of surface temperature (and thus thermal emission) is controlled by the patterns of incident stellar energy and surface albedo, both of which are symmetric about the substellar point. In all cold cases, even with global mean temperatures dipping into the low 200s K, a small circular patch of open-ocean is present at the substellar point, with temperatures marginally above freezing and few clouds (Figure 2, 4, 5). All cold cases have the emission maximum observed at secondary eclipse when the open-ocean patch on the day-side is in view of the observer, and the emission minimum at transit when the icy cold night-side is in view of the observer. For cold worlds, phase dependent thermal emission spectra are generally symmetric, with spectra at eastern and western quadrature being identical in Figure 8.

Temperate planets exhibit more complex changes to their phase dependent emission spectra. Still, the water vapor window region provides the optimal bandpass to observe variations in planetary emission spectra as a function of phase. For slow rotating planets ($T_{eff} \geq 3300$ K) the emission maximum is observed at transit, because high ice clouds over the substellar hemisphere effectively trap thermal radiation, while thermal radiation readily escape to space through the clear-sky conditions on the night hemisphere (e.g. Yang et al. 2013; Haqq-Misra et al. 2018). Temperate slow rotators exhibit remarkable symmetry, with thermal emission at eastern and western quadrature being identical and overlapping nearly perfectly in Figure 8. For Rhines rotators ($T_{eff}$ = 3000 K and 3300 K), day-night differences in thermal emission spectra become muted. Differences, even in the water vapor window region, are small due to advection of both heat and clouds around the planet (Kopparapu et al. 2016; Kopparapu et al. 2017). Finally, for rapid rotators, maximum thermal emission through the water vapor window region is observed at secondary eclipse, followed closely by the emission observed at western quadrature. For fast rotators, clouds are strongly advected eastward, completely off of the substellar point into the eastern hemisphere. Emission from the warm planet surface, located immediately under and west of the substellar point, is able to escape directly to space. Ice cloud layers form only over the eastern terminator, reducing outgoing longwave radiation in those regions (Haqq-Misra et al. 2018).

Incipient runaway cases tend to yield an emission maximum in the water vapor window region observed at secondary eclipse, due to intense stellar absorption by near-infrared water vapor absorption bands, strong middle atmosphere heating, and subsequent thermal emission to space. However, some variability in this trend is observed due to patchy and inconsistent clouds, probably due to the short duration and short temporal averaging period for these simulations. More strikingly incipient runaway cases may be discerned by the obvious thermal emission maximum also found in the ~4 μm water vapor window region. This result is qualitatively similar to the results of Leconte et al. (2013) for runaway greenhouse states on highly irradiated water-limited land planets, and the 1D line-by-line calculations of Goldblatt et al. (2013). However, Goldblatt et al. (2013) predicts that thermal emission near ~4 μm is only possible when a planet reaches the terminal state of the runaway, with surface temperatures of ~1800 K coincident with the emitting layer of the atmosphere becomes hot enough to emit at ~4 μm. Note the ~4 μm emission peak shown in Goldblatt et al. (2013) is significantly larger than those found in our study. Goldblatt et al. (2013) assumed a saturated atmosphere. Both here and in the Leconte et al. (2013), the runaway cases simulated remain subsaturated despite atmosphere temperatures beyond ~400 K and large specific humidities. Presumably, if not for numerical limitations of our model, our incipient runaway greenhouse cases would continue to warm until excessively hot steam atmospheres are obtained, as depicted with 1D radiative transfer modeling of Goldblatt et al. (2013). As such, the thermal emission feature shown here at ~4 μm should continue to strengthen as the runaway process continues. However, the subsaturation of the atmosphere means that the ~4 μm feature may appear very earlier in the temporal evolution of a runaway greenhouse state, and also possibly for other warm water-rich worlds. Perhaps more pertinently, the emergence of a modest ~4 μm emission feature may allow us to discern between planets with surface temperatures of 300 K and ~400 K, providing an important observable constraint on planetary habitability.

### 5.2. Albedo Spectra

In Figure 9, we show phase dependent spectrally resolved albedos for cold, temperate, and **incipient** runaway cases (columns), around each of our 6 stars (rows).

Figure 9 is arranged and color coded similarly to Figure 8, except that the albedo at transit is by definition undefined and thus omitted. Significant differences are found in the reflectance spectrum between climate states. Spectra are shown for 3 phases; western quadrature (gold), secondary eclipse (red), and eastern quadrature (light blue). In general, water vapor absorption features manifest as low-albedo regions of the albedo spectrum. Near-infrared water vapor window regions are controlled by a combination of cloud and surface reflectivities, and manifest as maxima in the albedo spectrum. Water vapor absorption is weak at visible wavelengths, thus visible albedos are also controlled by cloud and surface albedos, as well as Rayleigh scattering.

For cold climates, reflection off of the surface dominates the albedo spectrum, because these atmospheres have relatively few clouds and sparse water vapor. In our model, there is a step-wise division between the visible and near-infrared snow and ice albedos at ~0.7 μm. This division is clearly evident in phase dependent albedo spectra, indicating that the surface dominates the albedo for these cold and relatively dry and cloud free worlds. For slow rotating ($T_{eff} \geq 3300$ K) cold climates, there is a high degree of symmetry in reflected light phase curves. At visible wavelengths, reflection is always greatest at eastern and western quadrature and tend to overlap on the plot, while the albedo at secondary eclipse is slightly lower. This is due to the presence of a small open-ocean spot on the substellar hemisphere, which increases the disk averages albedo, compared to the completely ice-covered terminator regions. For cold climates that are faster rotating ($T_{eff} \leq 3000$ K) we observe different patterns. At visible wavelengths, the albedo at eastern quadrature is greatest, while the albedo at western quadrature and at secondary eclipse are less respectively. For these planets, increased zonal flow advects the sparse clouds eastward, and also results in enhanced snow fall over sea ice in the eastern hemisphere, leading to slightly higher surface albedos compared to the western hemisphere. For cold climates, near-infrared water vapor absorption features are discernable but weak. Some phase dependent variability is observed in the near-infrared window regions, generally mimicking the trends seen at visible wavelengths.

For temperate climates, visible albedos are controlled by clouds rather than the surface albedo, and near-infrared water vapor absorption features show prominently in the albedo spectrum. Recall that temperate climates have thick cloud cover over the substellar hemisphere. For slow rotating temperate worlds, eastern quadrature and secondary eclipse phases have about the same visible wavelength albedo, while the albedo is slightly lower at western quadrature. Even for slow rotators, the subtle effect of a small Coriolis parameter exists, preferentially favoring cloudiness in the eastern hemisphere over the western hemisphere (Figure 4). This trend holds true for planets around all stars with $T_{eff} \geq 3000$ K. For rapidly rotating temperate worlds, the strong advection of clouds eastward means that the eastern quadrature has a substantially higher visible albedo than is seen at secondary eclipse or western quadrature. For all temperate climates, near-infrared water vapor absorption bands show prominently. This is due to the combined effect of thick substellar clouds which raise albedos in water vapor window regions, and enhanced upper atmosphere water vapor amounts (e.g. Kopparapu et al. 2017; Fujii et al. 2017) which increase absorption within water bands. The resolution of the near-infrared water-vapor bands in reflectance spectra may be an important signpost for habitable worlds.

Incipient runaway cases manifest similarly around each stellar type. In these hot moist climates, there are few clouds on the irradiated hemisphere, large water-vapor specific humidities, and no sea ice. Ocean albedos are low, thus reflected light is dominated by Rayleigh scattering. The spectral albedo longward of 0.7 μm is generally less than ~0.2. Shortward of 0.7 μm reveals a very steep Rayleigh slope, as water is only very weakly absorbing, but enhanced specific humidities contribute to increased scattering.

### 5.3. Secondary Eclipse Spectra

To get a clearer comparison between the spectra of different climate states, here we collect thermal emission and albedo secondary eclipse spectra from each climate state, around each of our 6 stars (Figure 10). Generally, in the thermal emission (Figure 10, left column), the hotter the climate state, the more emission from the planet. This is true at all wavelengths, but is particularly evident in the water-vapor window region of 8 – 13 μm. Subsaturation of the atmospheres prevents the water vapor window region from closing, preventing the emergence of a hard radiation limit for moist atmospheres (Goldblatt et al. 2013). As noted earlier, for incipient runaway cases strong emission becomes evident in the ~4 μm water vapor window region. The ~4 μm feature may be a signpost for a planet undergoing a runaway greenhouse (see also Goldblatt et al. 2013; Leconte et al. 2013). However, based on secondary eclipse spectra alone, it may be difficult to discern between cold, cool, and temperate climate states. To discern these climate states, one also needs to additionally consider variations in phase dependent thermal emission spectra (Figure 8), and also planetary albedos.

For phase dependent albedos (Figure 9, right column), as noted earlier, cool and temperate climates have the highest albedos in both the visible and near-infrared wavelengths due to the omnipresent thick substellar cloud deck for slow and Rhines circulation regimes. For slow rotators ($T_{eff} \geq$ 3300 K) spectral albedos for cool and temperate climates are nearly indistinguishable, with high albedos in the visible and prominent water vapor absorption features evident in the near-infrared. Cold climates, have a lower visible albedo than temperate worlds, and also have muted water vapor features in the near-infrared. Incipient runway cases have the lowest spectral albedos except at ~0.2 μm, the shortwave limit of our model, where atmospheric scattering is significant. For rapid rotating cases ($T_{eff}$ = 2600 K), the albedo is inversely related to the planet temperature, evident at all wavelengths. Rhines rotation cases $T_{eff}$ = 3000 K) represent a transition between slow and rapid rotating regimes.

In an effort to constrain observability, in Figure 11 we show the combined light secondary eclipse spectra for each climate state, around each of our 6 stars. Spectra are given in planet/star contrast in units of parts-per-million (ppm). The combined light spectra shows the reflected stellar light plus the thermal emitted light from the planet in a single spectrum, here plotted relative to the emitted light from the star. Generally, the combined light spectral signals measured at secondary eclipse from any climate state only exceeds ~1 ppm at wavelengths longer than ~12 μm for planets around 4500 K stars, and ~6 μm for planets around 2600 K stars. Lower effective temperature stars are smaller, and thus dimmer, meaning that planet spectral signals are relatively larger. Part-per-billion (ppb) precision would allow us to spectrally resolve all thermal (≥4 μm) water vapor and cloud features for planets around M-dwarf stars. Reflected light at wavelengths less than ~2 μm, are also generally amenable to detection at ppb precision in all cases, except for runaway greenhouse climates around 4500 K stars. The most difficult spectral region to detect lies between 2 – 3 μm, where part-per-trillion (ppt) precision would be required to resolve planet spectral features, except around the smallest M-dwarf stars where ppb may suffice. While ppb and ppt precisions are out of bounds currently for observations of combined light secondary eclipse spectra, our calculations here demonstrate that climate states may be discriminated based on the spectral signals resulting from expected water vapor and cloud distributions. Future missions and instruments should aim for ppb precision in order to characterize terrestrial planets around M-dwarf stars in thermal emission and reflected stellar light.

## 6. Variation Spectra

Selsis et al. (2011) coined the term "variation spectrum" as the amplitude of variation in the radiance emitted or reflected by an exoplanet over the course of its orbit. In practice, the variation spectrum is calculated as the maximum observed radiance minus the minimum observed radiance, measured over the planet's orbit. In Figure 12, we plot the combined light variation spectra for incipient runaway (red), temperate (gold), cool (light blue), and cold (dark blue) climates, around each of our 6 stars. The left column shows the variation spectra in units of radiance (W $m^{-2}$ $\mu m^{-1}$) measured at the top of the atmosphere, while the right column shows the variation spectra in planet/star contrast in units of ppm. Similar to Figure 11, combined light is considered as the reflected stellar plus the outgoing thermal emitted radiation from the planet.

Spectral differences in the variation spectra can yield information about climate states. In radiance units, the variation spectra at visible wavelengths is large. For shortwave variation spectra, the maximum reflection is observed at secondary eclipse, while the minimum in reflection occurs by transit where by definition no light is reflected towards to observer. Thus, the shortwave the variation spectrum indicates the reflectivity of the substellar hemisphere. The absolute maximum of the variation spectra in radiance units corresponds with the wavelength of peak emission from the star (Figure 1). As noted previously, temperate and cool climate states generally reflect more light than do cold and icy climates because water clouds have higher broadband albedos than does sea ice, when irradiated by M-dwarf stars (Shields et al. 2013).

Around each star and for all climate states, a local minimum in the variation spectra is found corresponding with the ~2.7 μm water vapor absorption band. The magnitude of the variation spectra at 2.7 μm is inversely related to the planet temperature, with cold planets having the most variation, and incipient runaway states having the least. In the ~4 μm water vapor window, runaway greenhouse climates exhibit a local maximum that is not found for other climate states. Other climate states tend to exhibit either a modest minimum in the variation spectra or remain flat at ~4 μm, due to it residing at the shoulder of both thermal and stellar radiative fields. Generally, a local minimum is also observed near the ~6 μm water vapor feature, which is particularly sharp for incipient runaway cases. All climates around all stars exhibit maximum in the thermal variation spectra in the water vapor window region.

Figure 12 right column shows the combined light variation spectra of planets relative to the stellar brightness. The brightness of the stars relative to planets means that the planet/star contrast at short wavelengths is significantly lower than that observed at thermal wavelengths. Still, many of the same trends hold true that were noted for variation spectra in radiance units. Local minima are seen in water vapor absorption bands, most notably at ~2.7 μm and ~6 μm. At ~4 μm, runaway greenhouse states exhibit a maximum, while all other climate states display either a modest minimum or flat spectra. All climates exhibit their largest variation at thermal wavelengths, where the stellar emission is weak. Cold climates exhibit the largest variation spectra beyond ~10 μm, while that from runaway states is muted.

The precision required to observe variation spectra is greater than that required to measure secondary eclipse spectra (Figure 11), because phase dependent variations in radiance are, naturally, less than their magnitude. To capture variation spectra at wavelengths ≥6 μm, ppb precision is generally sufficient for planets around all stars studied. However, some of the most interesting and tell-tale spectral signals reside in the 2 – 6 μm range, which becomes increasingly challenging to observe from planets around large stars.

## 7. Discussions

In this section we discuss numerous caveats to the work presented here. In this study, we have only considered Earth-sized aquaplanets, with 1 bar of $N_2$ along with $H_2O$ and clouds. The presence of different absorbing species would affect climate, phase curves, and phase dependent spectra. Naturally, the addition of other greenhouse gases would result in warmer climate states for a given stellar flux. Also, gases such as $CO_2$ and $CH_4$ are generally well-mixed species, meaning that they are evenly distributed about the planet, and thus may uniformly reduce the magnitude of broadband thermal emission phase curves. While well-mixed greenhouse gases would surely be evident in spectra, it is unlikely that they would contribute significantly to the variation spectra (i.e. phase dependent variability). For instance, Boutle et al. (2017) showed, based on tidally-locked simulations of Proxima b, that phase dependent emission spectra are insensitive to phase variations in the 15 μm $CO_2$ absorption band. A similar conclusion was also reached by Turbet et al. (2016) based on Proxima b simulations with a different model. Clouds dominate phase dependent variability, not gaseous absorbing species, and thus phase dependent variability is most evident in window regions of the spectra. Still, strong absorption by $CO_2$ at 15 μm or by $O_3$ at 9.6 μm, could encroach upon the water-vapor window region, and partially obscure the variation spectra that have been presented here. Additionally, planets with drastically different atmospheric compositions and total surface pressures would be expected to have significant differences in temperature, atmospheric circulation, heat transport, lapse rates, convection, and cloud formation. Each of these aspects would affect thermal and reflected phase curves morphologies and amplitudes.

Analysis of *Keple*r planets indicates that terrestrial planets could feasibly be as large as 1.5 Earth radii, with larger planets being Neptune-like instead of terrestrial (Rogers, 2015; Wolfgang et al. 2016). Changes in planetary radius could have modest effects on atmospheric dynamics and circulation regimes, quantified by changes to the Rossby radius of deformation vis-à-vis the planet radius (Carone et al. 2018). Changes in planet size and internal composition would also be expected to change the surface gravity. Changes in surface gravity would affect many aspects of the climate system, including the lapse rate, scale heights, and particle sedimentation rates, all of which could strongly affect the nature of clouds and climate on extrasolar planets. Naturally, changes to clouds and climate would feedback upon the phase dependent thermal and reflected light signals. Larger sized planets would be more favorable to observation, yielding larger planet/star contrasts than are shown here, and vice-versa. We have also assumed that these planets are all tidally locked into synchronous rotation with their host star. Resonant rotation states and eccentric orbits would also influence phase dependent observations (e.g. Wang et al. 2014; Bolmont et al. 2016), which we have not considered here.

Here, for simplicity, computational convenience, and continuity with past works (Kopparapu et al. 2017; Haqq-Misra et al. 2018), we have assumed a slab ocean with zero ocean heat transport. Ocean heat transport may have a significant effect on climate and also the location of clouds (Way et al. 2018**).** Dynamic ocean heat transport on completely ocean covered worlds (i.e. no continents), leads to warmer global mean temperatures and a reduction in day-night temperature differences on synchronously rotating planets (Hu & Yang 2014; Del Genio et al. 2019). Yang et al. (2019) argue that ocean heat transport is critical for the treatment of cold, partially ice-covered planets around M-dwarf stars, however it does not have a meaningful effect on the climate and thermal emission phase curves for warm terrestrial planets ($T_s \sim 300$ K) residing close to the inner edge of the habitable zone. The presence of continents also presents a significant uncertainty in the net effect of ocean heat transport on climate, as continents can reroute or eliminate day-to-night ocean heat transport entirely, resulting in climates states that are similar to those found in simulations with zero ocean heat transport (Del Genio et al. 2019; Yang et al. 2019). In this work, we have argued that clouds are likely more important in modulating phase curves for habitable planets, compared to surface energy transports. Still, ocean heat transport patterns combined with variations in continental configurations and topographic prominences could have significant complex feedbacks on the surface temperature, sea-ice and snow cover, and cloud distributions respectively, all of which affect thermal and reflected light phase curves. In future works, our group plans to couple our GCM with a dynamic ocean model to study the interactions between oceans, continents, topography, and observable features. However, the long equilibration times for a coupled dynamic atmosphere-ocean models still presents a significant computational hinderance for conducting 3D simulations across wider parameter spaces, as we have done here.

The suite of simulations studied here, represents only a small fraction of the possible worlds that could reside in the habitable zones of M-dwarf stars. While here we have demonstrated how water vapor and clouds may affect thermal and reflected light phase curves for Earth-like aquaplanets, more work is needed to systematically map how other parameters would affect phase dependent observations. Future planned work will expand our modeling to include ocean heat transport, different atmospheric compositions, planet sizes, and surface configurations.

## 8. Conclusions

Terrestrial planets orbiting low-mass stars will provide our best opportunity for detecting and verifying a habitable world in the coming decades. Here, we suggest that broadband and spectrally resolved thermal emitted and reflected light phase curves can be used to characterize the climate states of tidally locked terrestrial extrasolar planets orbiting M-dwarf stars. The spatial distributions of water vapor, clouds, and surface ice significantly impact phase dependent spectral signals from terrestrial planets. In turn, the mean climate state of a planet controls the partitioning of water into vapor, cloud, and ice phases. Thus, by carefully studying the phase dependent signals of water in each of its thermodynamic phases, we may connect the dots between observations and climate states for terrestrial extrasolar planets around M-dwarf stars. Here, we have calculated theoretical broadband and spectrally resolved thermal emission and reflected light phase curves, as well as variation spectra, based on 3D climate modeling simulations of extrasolar planets in a variety of climate states, ranging from cold, ice dominated worlds with $T_s \sim 205$ K, to hot, moist, optically thick worlds with $T_s \sim 400$ K and with a runaway greenhouse initiated. Cold, temperate and incipient runaway greenhouse climate states exhibit unique phase dependent spectra in both thermal and reflected light. For potentially habitable worlds, we recommend probing the spectral window regions, where variability in clouds and surface properties will show most strongly. In particular, we recommend probing water vapor window regions in the near-infrared, 4 μm, and in the 8 – 13 μm region. Furthermore, we recommend that future missions and instruments aim for part-per-billion precisions, in order to sufficiently characterize water vapor and cloud features in the atmospheres of terrestrial planets.

**Acknowledgements**

R.K., E.T.W., and J.H.-M. acknowledge funding from NASA Habitable Worlds grant NNX16AB61G. R.K. and E.T.W. acknowledge funding from the Sellers Exoplanet Environments Collaboration. R.K., E.T.W., and J.H.-M. acknowledge funding from the Virtual Planetary Laboratory under Cooperative Agreement number NNA13AA93A. E.T.W. and R.K. acknowledge NASA Astrobiology Institute CAN7 award NNH13DA017C through participation in the Nexus for Exoplanet System Science. The 3D simulation outputs used in this work have been publicly archived, and are available at the following address: https://archive.org/details/SimulatedPhaseDependentSpectraOfTerrestrialAquaplanetsInMdwarfSystems, or by contacting the lead author, eric.wolf@colorado.edu.

## Figure Captions

**Figure 1:** Stellar spectra used in this study at the binned resolution of our model (solid lines.) The **solar** spectra is shown for comparative purposes (dotted line).

**Figure 2:** Maps of surface temperature from cold, cool, temperate, and incipient runaway greenhouse worlds (columns), around each of our 6 stars (rows). The panels are centered on the substellar point. The white line indicates the extent of sea-ice in each simulation. Note the range in orbital periods and incident stellar fluxes are indicated in the left-hand column, with the longest (lowest) orbital period (incident stellar flux) corresponding with the cold simulations, and the shortest (highest) orbital period (incident stellar flux) corresponding with the incipient runaway greenhouse.

**Figure 3.** Maps of the water vapor column (kg $m^{-2}$) for various climate states around each star. Same format as Figure 2.

**Figure 4.** Maps of the liquid cloud water column (g $m^{-2}$) for various climate states around each star. Same format as Figure 2. Note in the colorbar that clouds are shown in white.

**Figure 5.** Maps of the ice cloud water column (g $m^{-2}$) for various climate states around each star. Same format as Figure 2. Note in the colorbar that clouds are shown in white.

**Figure 6:** Broadband thermal emission phase curves calculated for all simulations, around each of our 6 stars. The fluxes given are the top-of-atmosphere values, disk-averaged as a function of phase. Phase curves are color coded according to climate state. Solid gold lines indicate temperate climates, solid light blue lines indicate cool climates, and solid dark blue lines indicate cold climates, with surface temperature ranges given in the legend. Dashed lines indicate incipient runaway greenhouse cases, and are representative of hot, moist, optically thick atmospheres.

**Figure 7:** Broadband albedo phase curves calculated for all simulations, around each of our 6 stars. Same format and color coding as Figure 6.

**Figure 8:** Phase dependent thermal emission spectra for cold, temperate, and incipient runaway cases, around each of our 6 stars. Plotted are the disk-averaged spectra at transit, western quadrature, secondary eclipse, and eastern quadrature. Note, for all cold climates and for temperate climates around star with $T_{eff} \geq 3700$ K, phase curves appear symmetrical and thus emission spectra at eastern and western quadrature are identical and tend to overlap and thus are difficult to distinguish without close inspection.

**Figure 9:** Phase dependent albedo spectra for cold, temperate, and incipient runaway cases, around each of our 6 stars. Identical format and color coding as Figure 8. Plotted are the disk-averaged spectra at western quadrature, secondary eclipse, and eastern quadrature. Note, some overlaps between phase dependent reflectance spectra occur, and thus can be difficult to distinguish without close inspection.

**Figure 10:** Thermal emission (left column) and albedo (right column) secondary eclipse spectra for incipient runaway, temperate, cool, and cold climates cases, around each of our 6 stars.

**Figure 11:** Combined light spectra at secondary eclipse for cold, cool, temperate, and **incipient** runaway, climate states around each of our 6 stars. Combined light is the reflected stellar light plus the thermal emitted light. Spectra are given in units planet/star contrast in parts-per-million.

**Figure 12:** Combined light variation spectra for cold, cool, temperate, and incipient runaway, climate states, around each of our 6 stars. The right column is in units of top-of-atmosphere radiances, and the left column shows planet/star contrasts in part-per-million.

**Figure 1:**

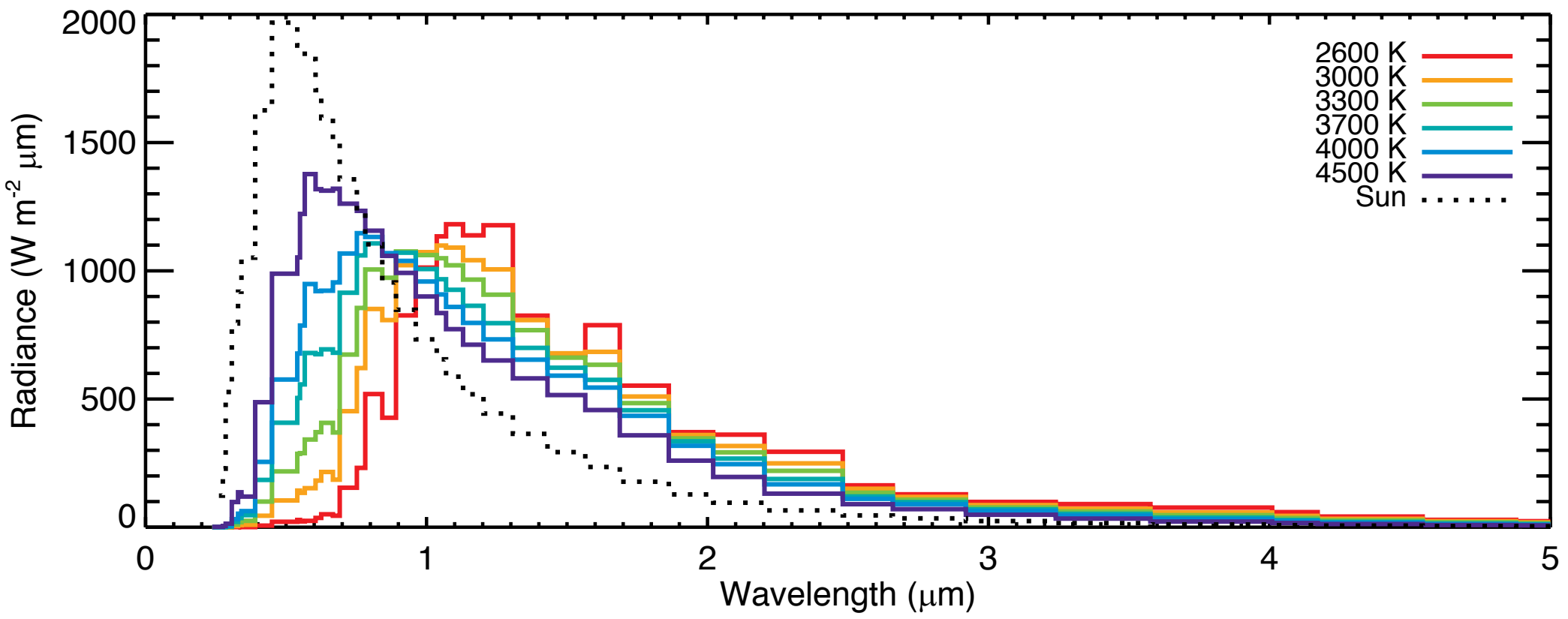

# Figure 2

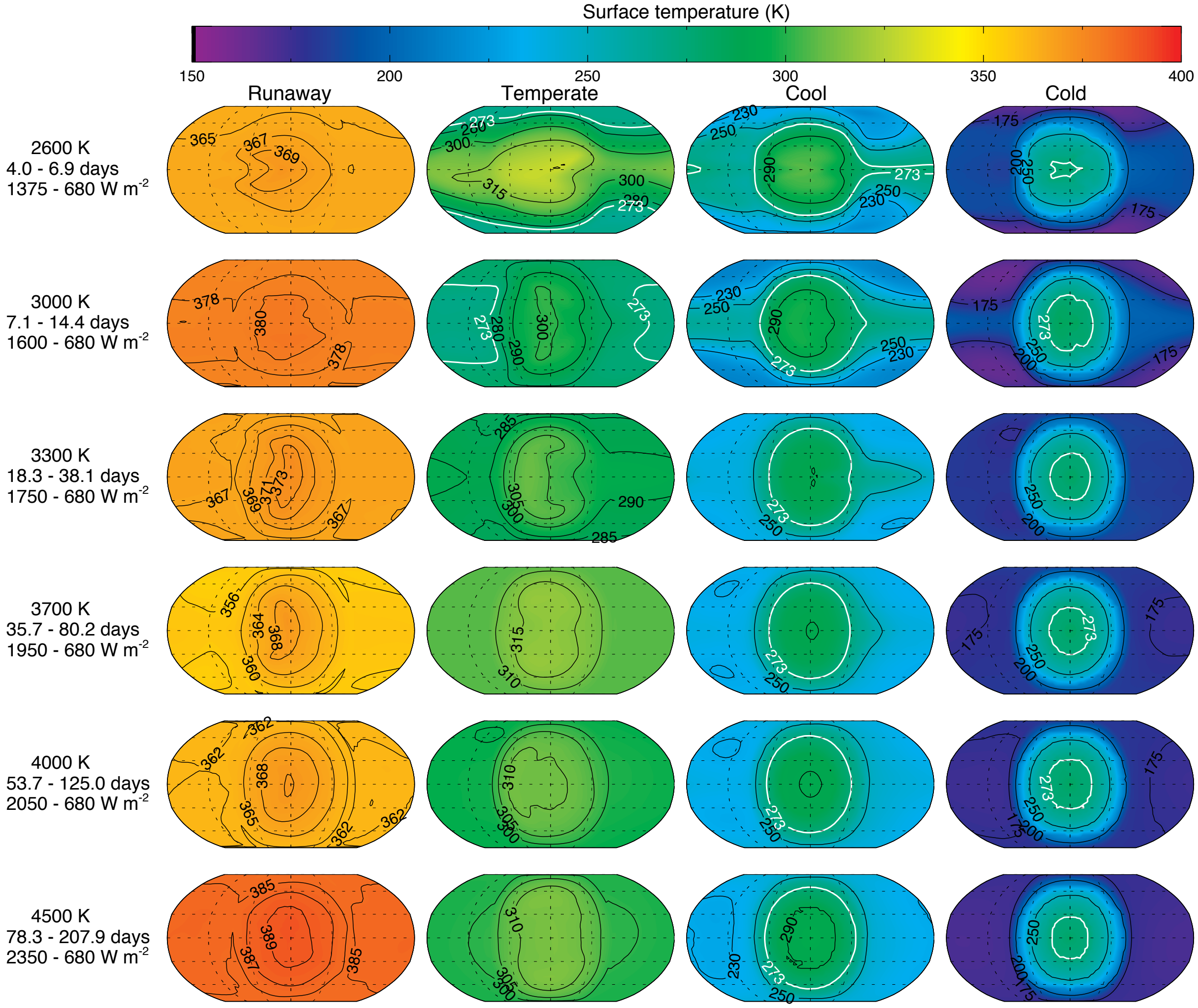

Surface temperature (K)
150
200
250
300
350
400
Runaway
Temperate
Cool
Cold
2600 K
4.0 - 6.9 days
1375 - 680 W m-2
3000 K
7.1 - 14.4 days
1600 - 680 W m-2
3300 K
18.3 - 38.1 days
1750 - 680 W m-2
3700 K
35.7 - 80.2 days
1950 - 680 W m-2
4000 K
53.7 - 125.0 days
2050 - 680 W m-2
4500 K
78.3 - 207.9 days
2350 - 680 W m-2

# Figure 3

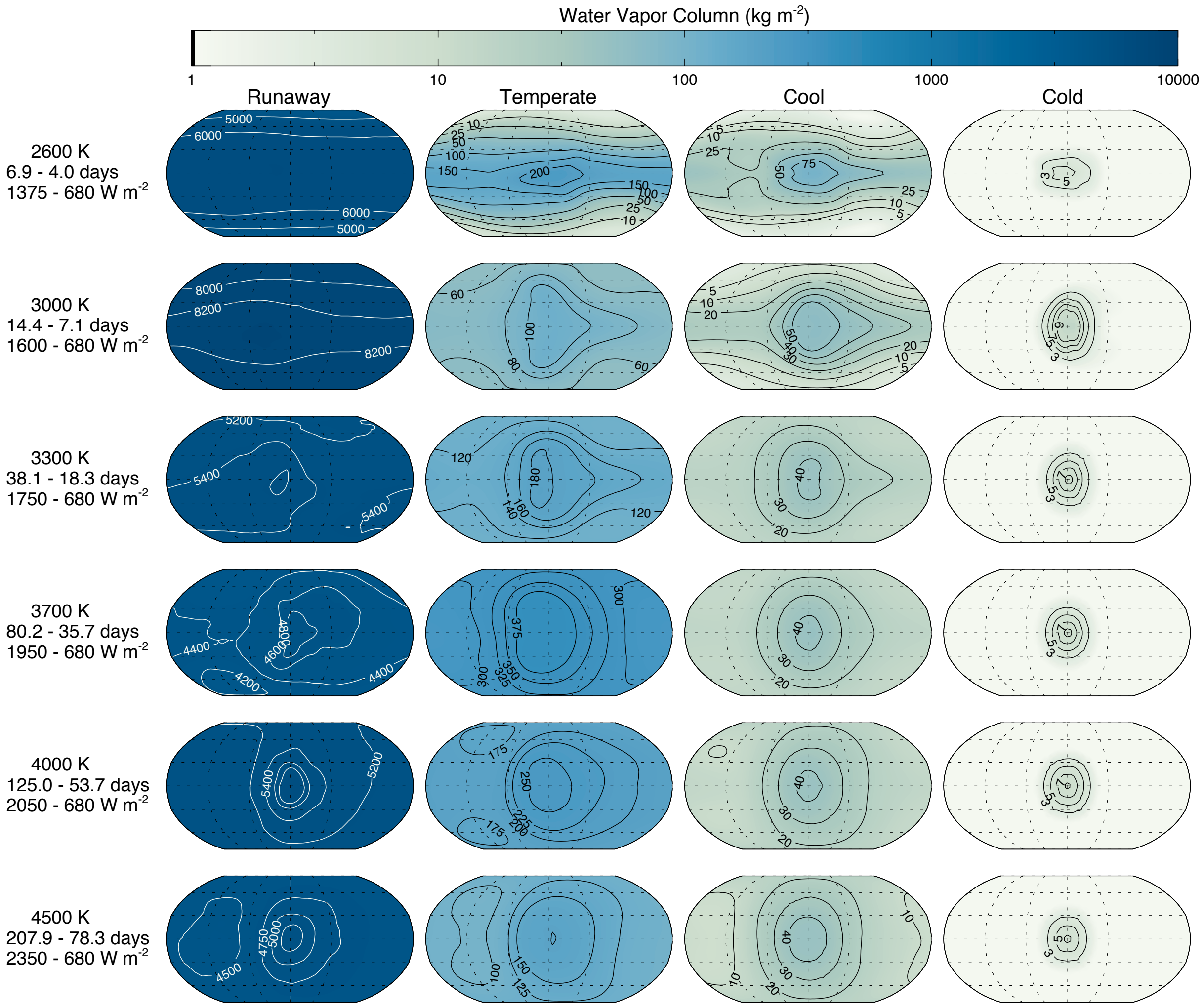

Water Vapor Column (kg m-2)
1
10
100
1000
10000
Runaway
Temperate
Cool
Cold
2600 K
6.9 - 4.0 days
1375 - 680 W m-2
3000 K
14.4 - 7.1 days
1600 - 680 W m-2
3300 K
38.1 - 18.3 days
1750 - 680 W m-2
3700 K
80.2 - 35.7 days
1950 - 680 W m-2
4000 K
125.0 - 53.7 days
2050 - 680 W m-2
4500 K
207.9 - 78.3 days
2350 - 680 W m-2

**Figure 4**

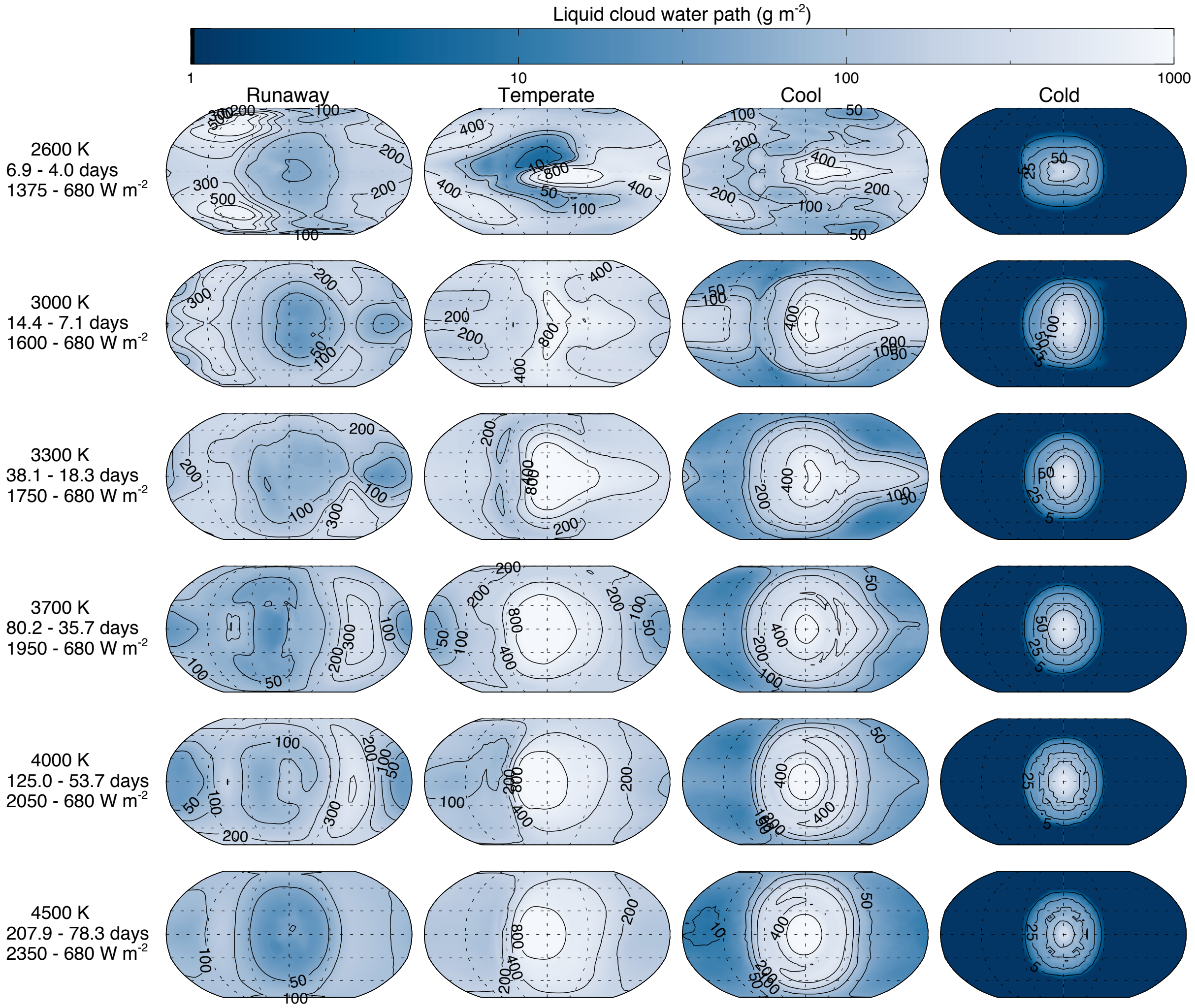
Liquid cloud water path (g m$^{-2}$)
1
10
100
1000
Runaway
Temperate
Cool
Cold
2600 K
6.9 - 4.0 days
1375 - 680 W m$^{-2}$
3000 K
14.4 - 7.1 days
1600 - 680 W m$^{-2}$
3300 K
38.1 - 18.3 days
1750 - 680 W m$^{-2}$
3700 K
80.2 - 35.7 days
1950 - 680 W m$^{-2}$
4000 K
125.0 - 53.7 days
2050 - 680 W m$^{-2}$
4500 K
207.9 - 78.3 days
2350 - 680 W m$^{-2}$

# Figure 5

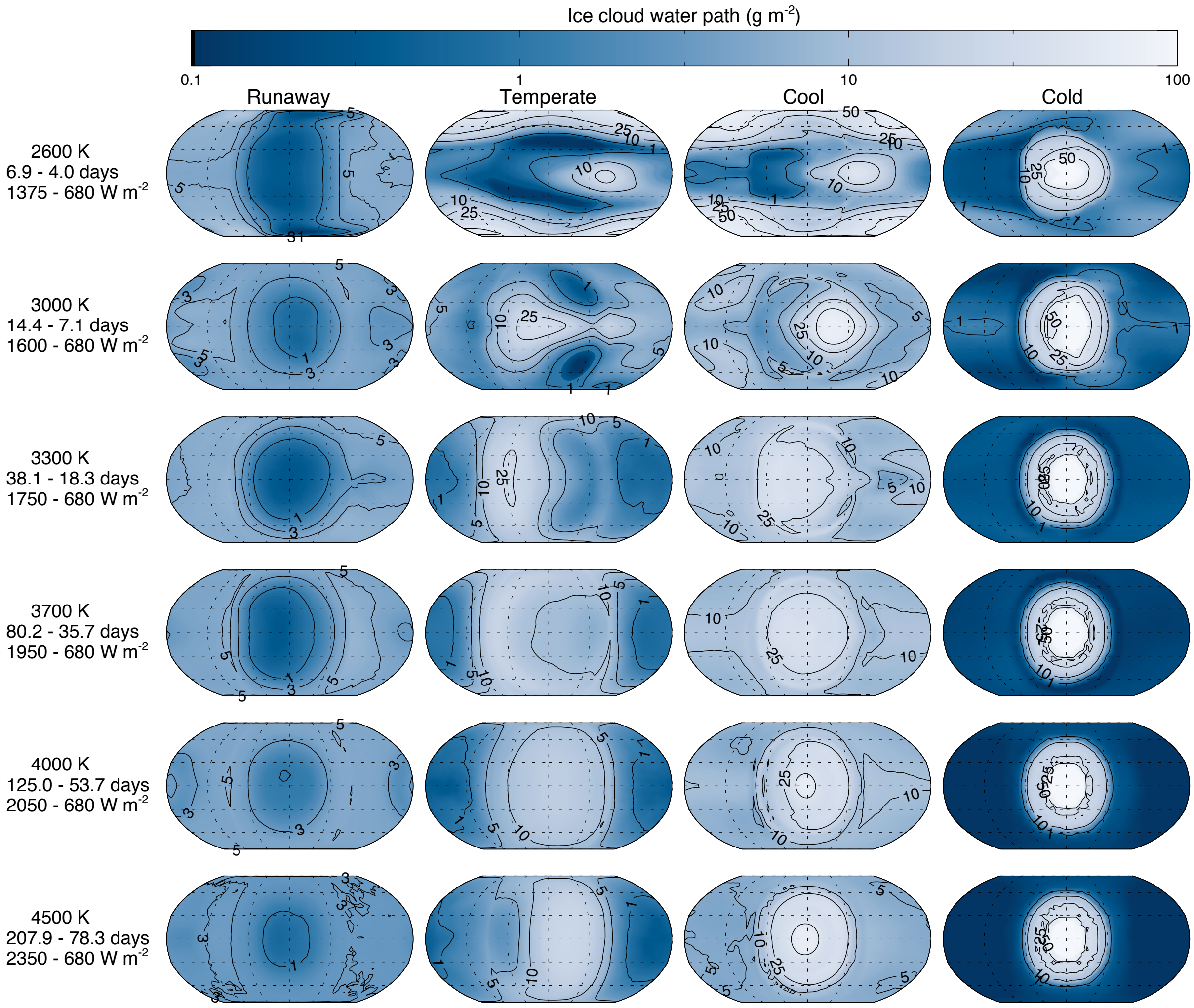

Ice cloud water path (g m$^{-2}$)
0.1
1
10
100
Runaway
Temperate
Cool
Cold
2600 K
6.9 - 4.0 days
1375 - 680 W m$^{-2}$
3000 K
14.4 - 7.1 days
1600 - 680 W m$^{-2}$
3300 K
38.1 - 18.3 days
1750 - 680 W m$^{-2}$
3700 K
80.2 - 35.7 days
1950 - 680 W m$^{-2}$
4000 K
125.0 - 53.7 days
2050 - 680 W m$^{-2}$
4500 K
207.9 - 78.3 days
2350 - 680 W m$^{-2}$

**Figure 6:**

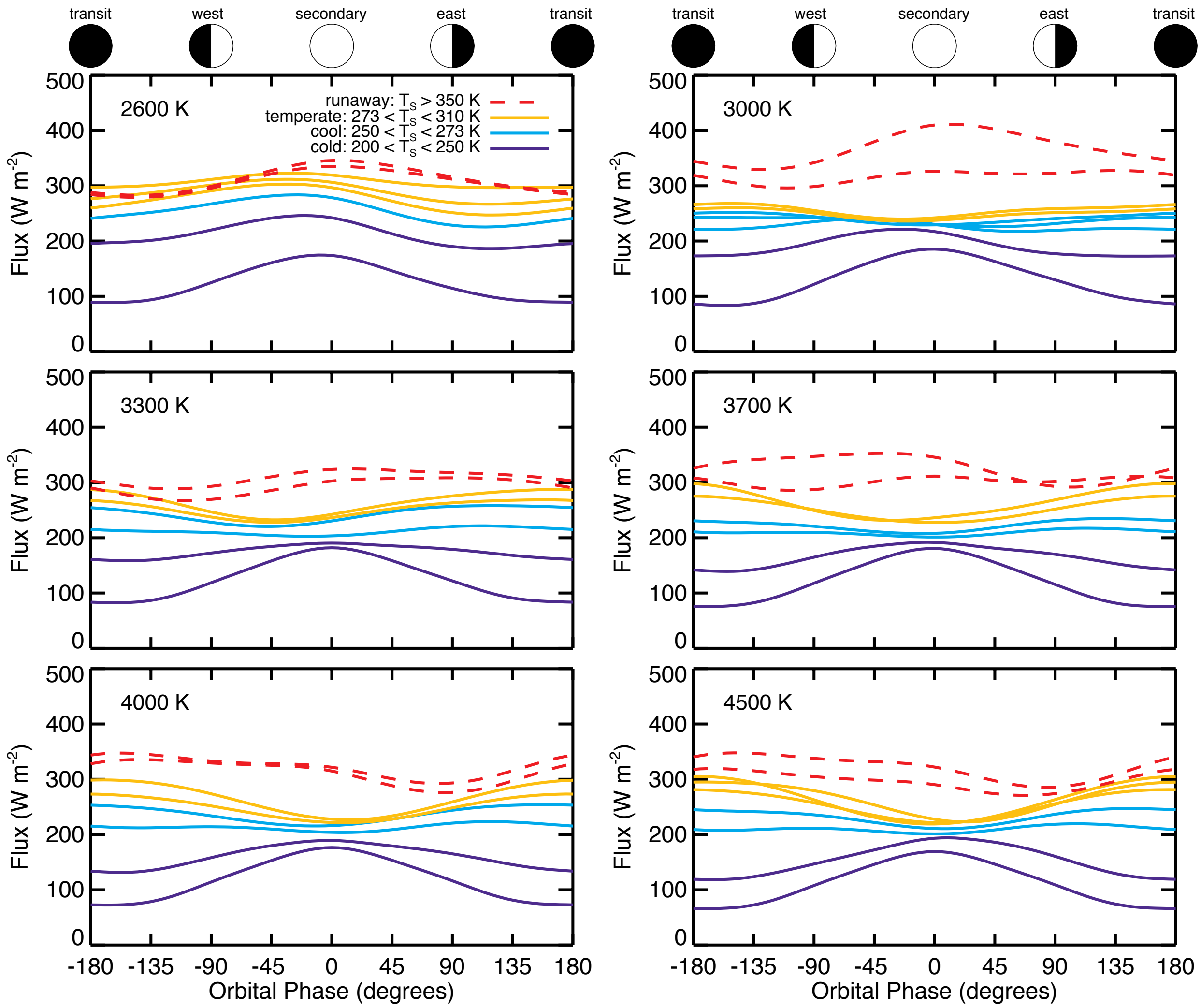

**Figure 7:**

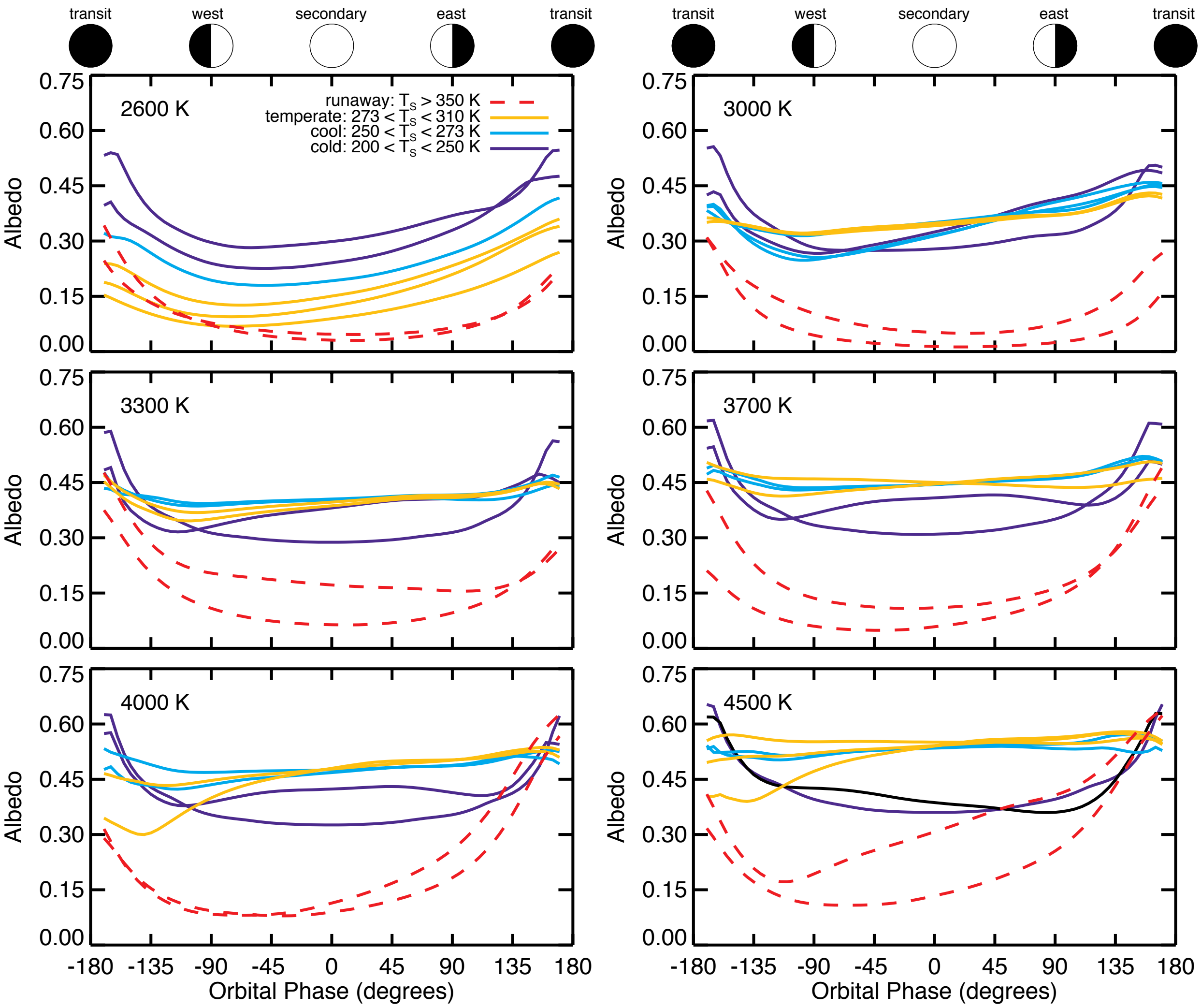

transit
west
secondary
east
transit
2600 K
3000 K
3300 K
3700 K
4000 K
4500 K
runaway: $T_S$ > 350 K
temperate: 273 < $T_S$ < 310 K
cool: 250 < $T_S$ < 273 K
cold: 200 < $T_S$ < 250 K
Albedo
Orbital Phase (degrees)

**Figure 8:**

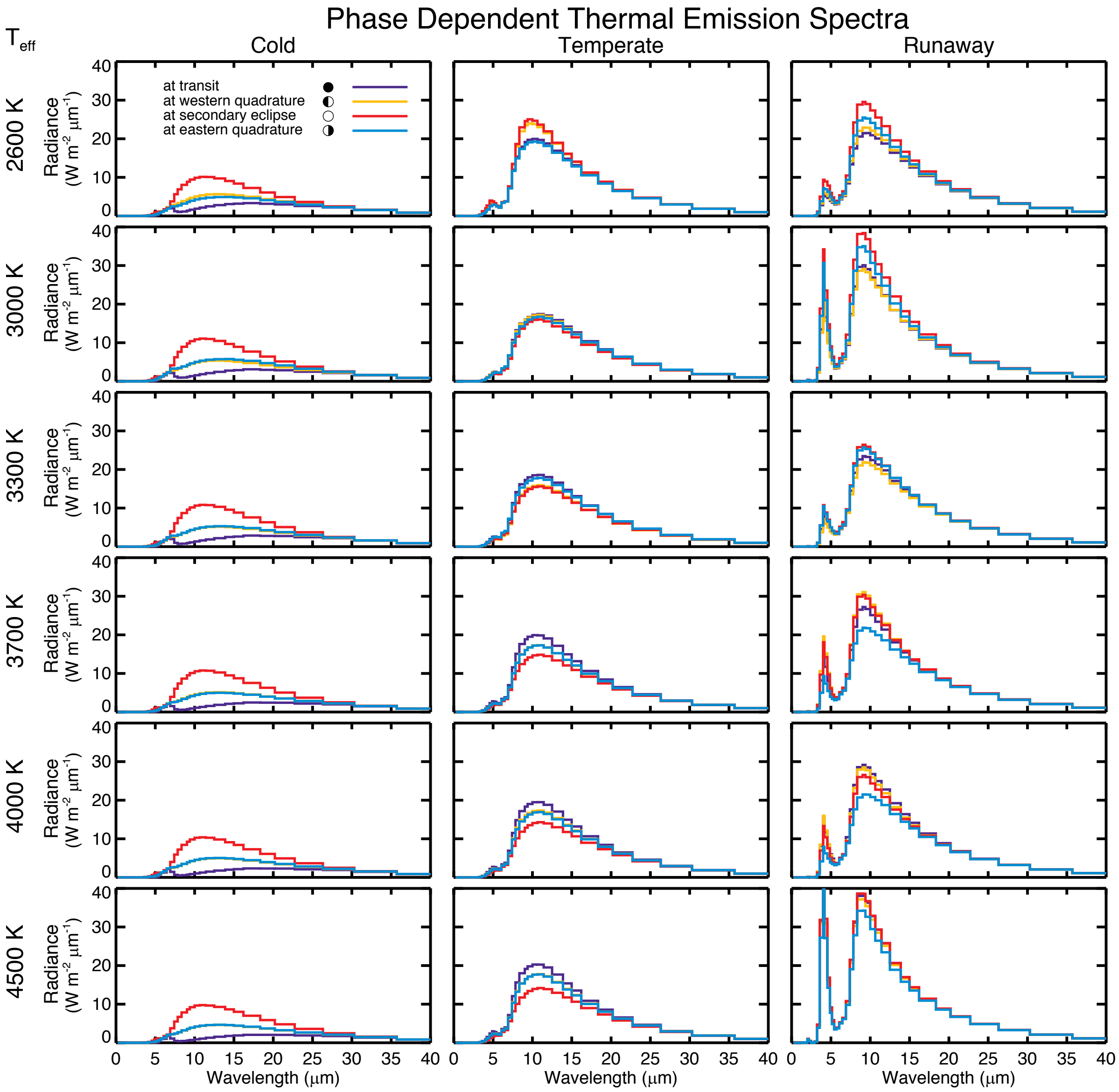

**Figure 9:**

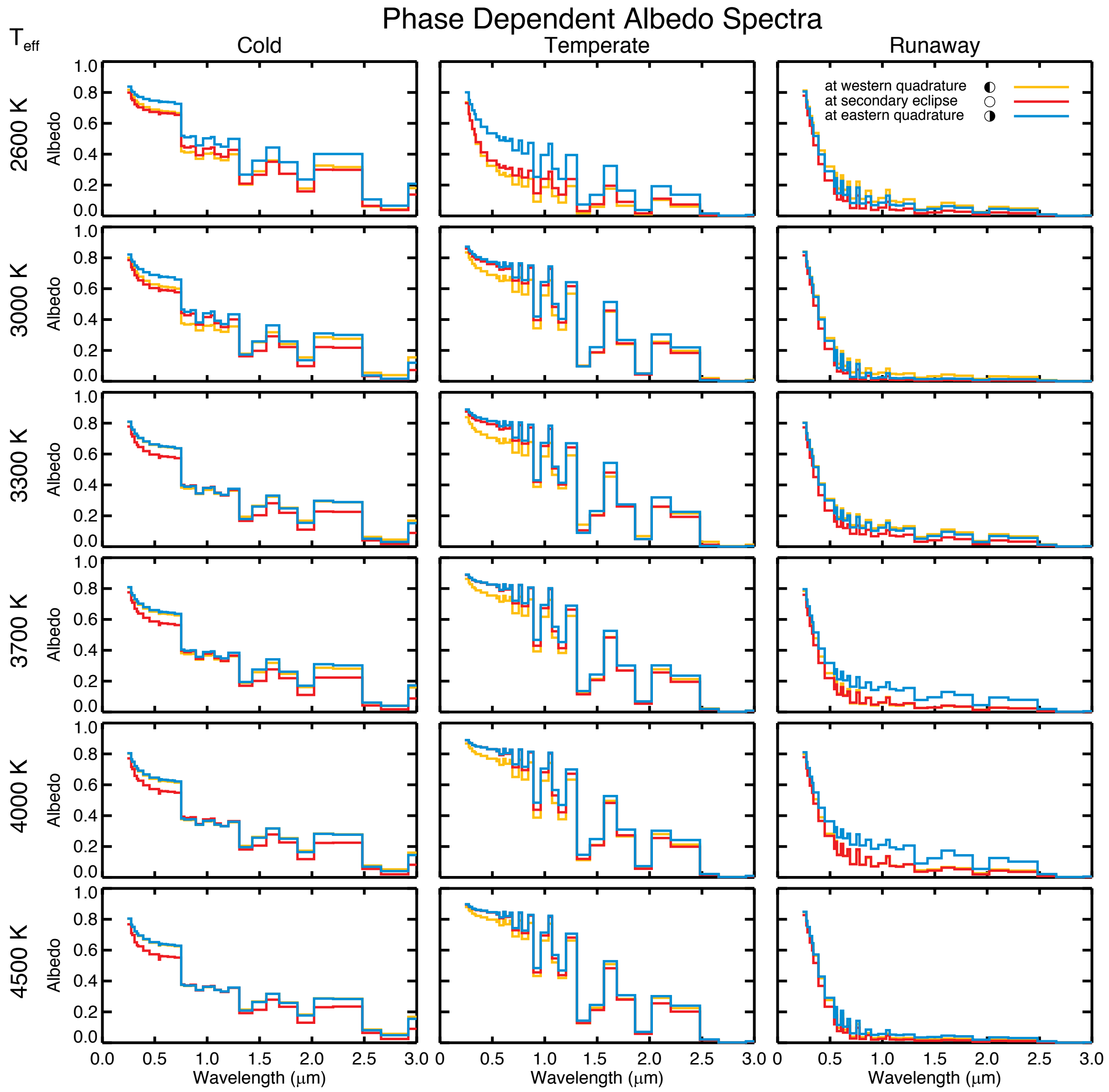

**Figure 10:**

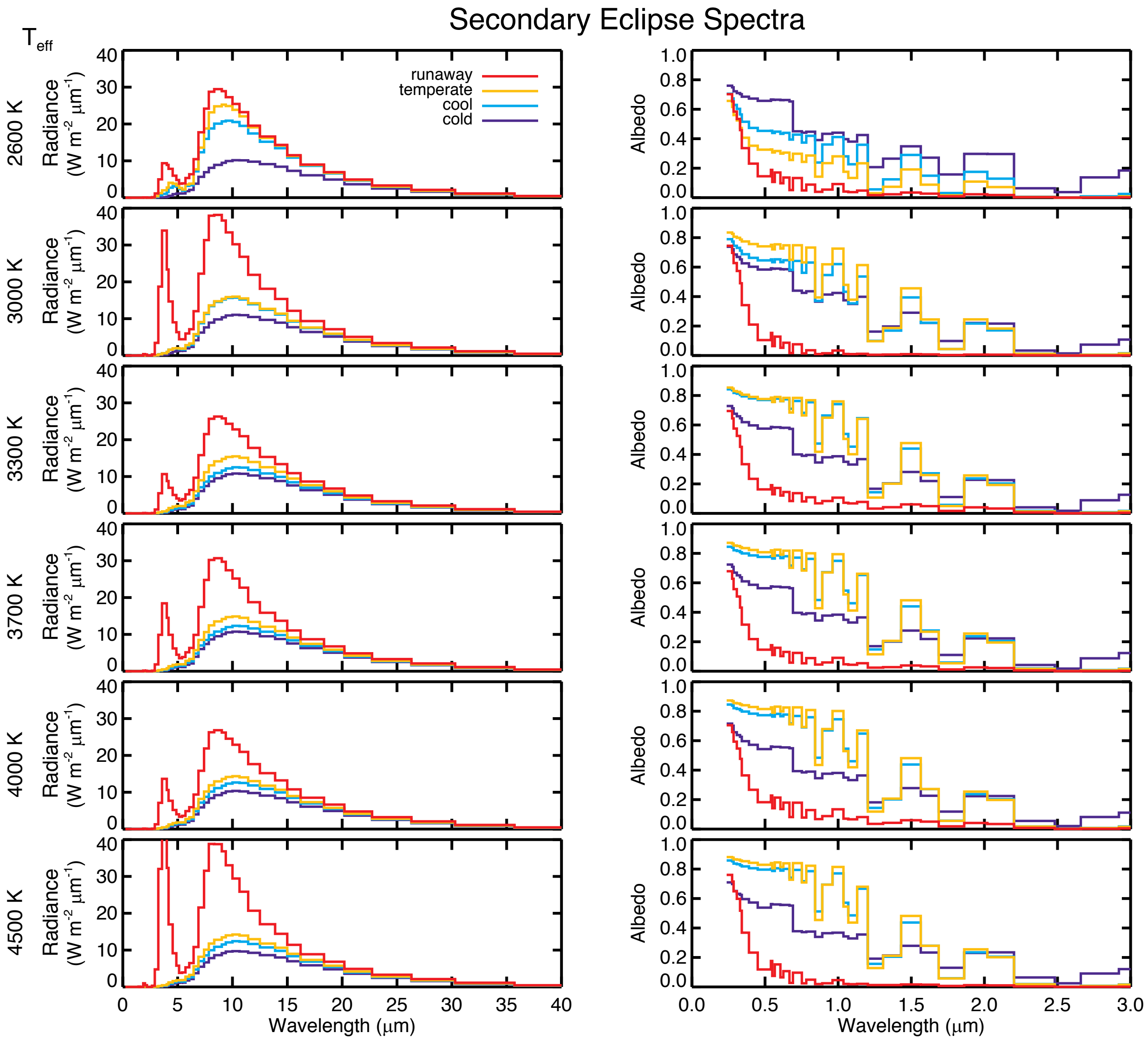

**Figure 11:**

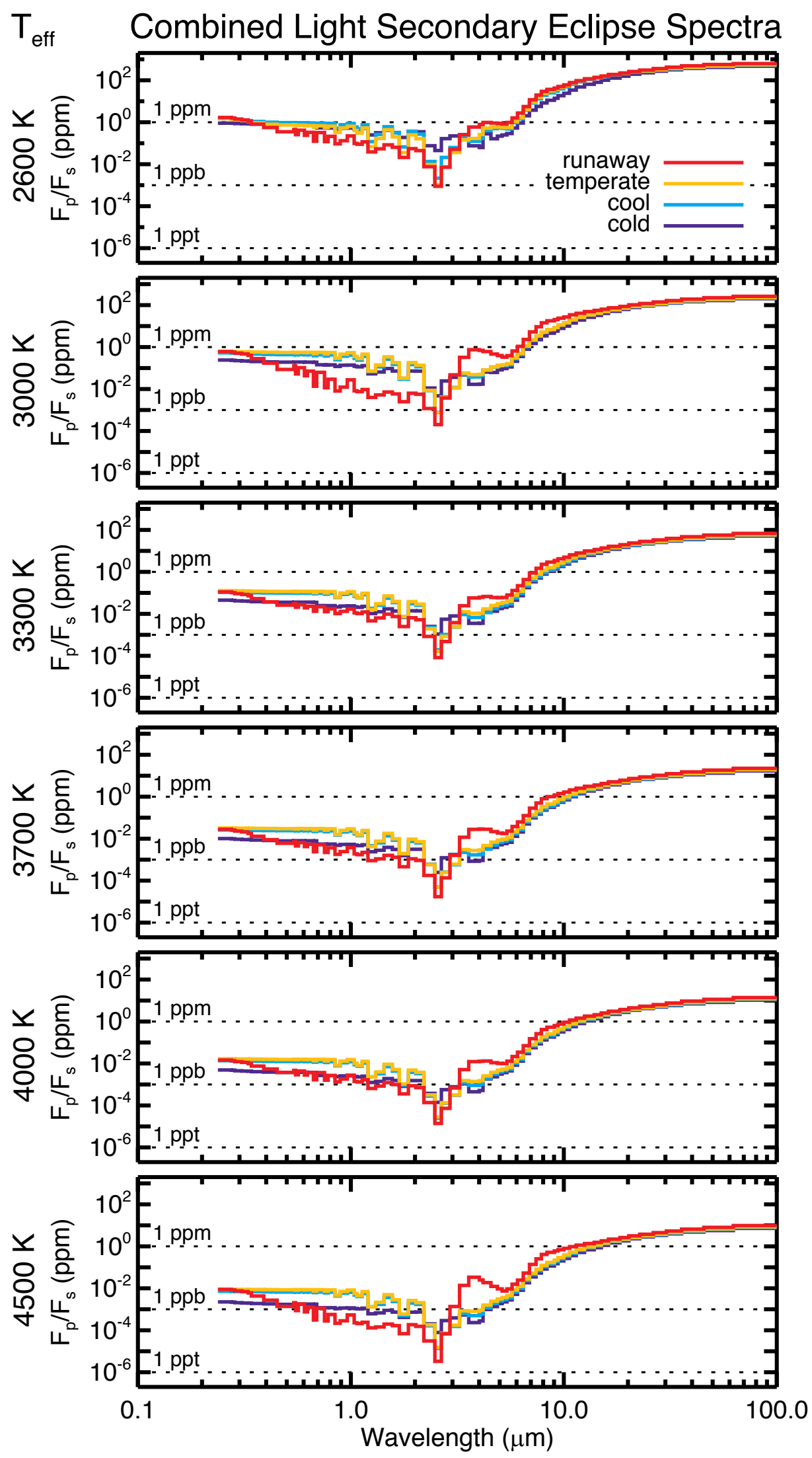

**Figure 12:**

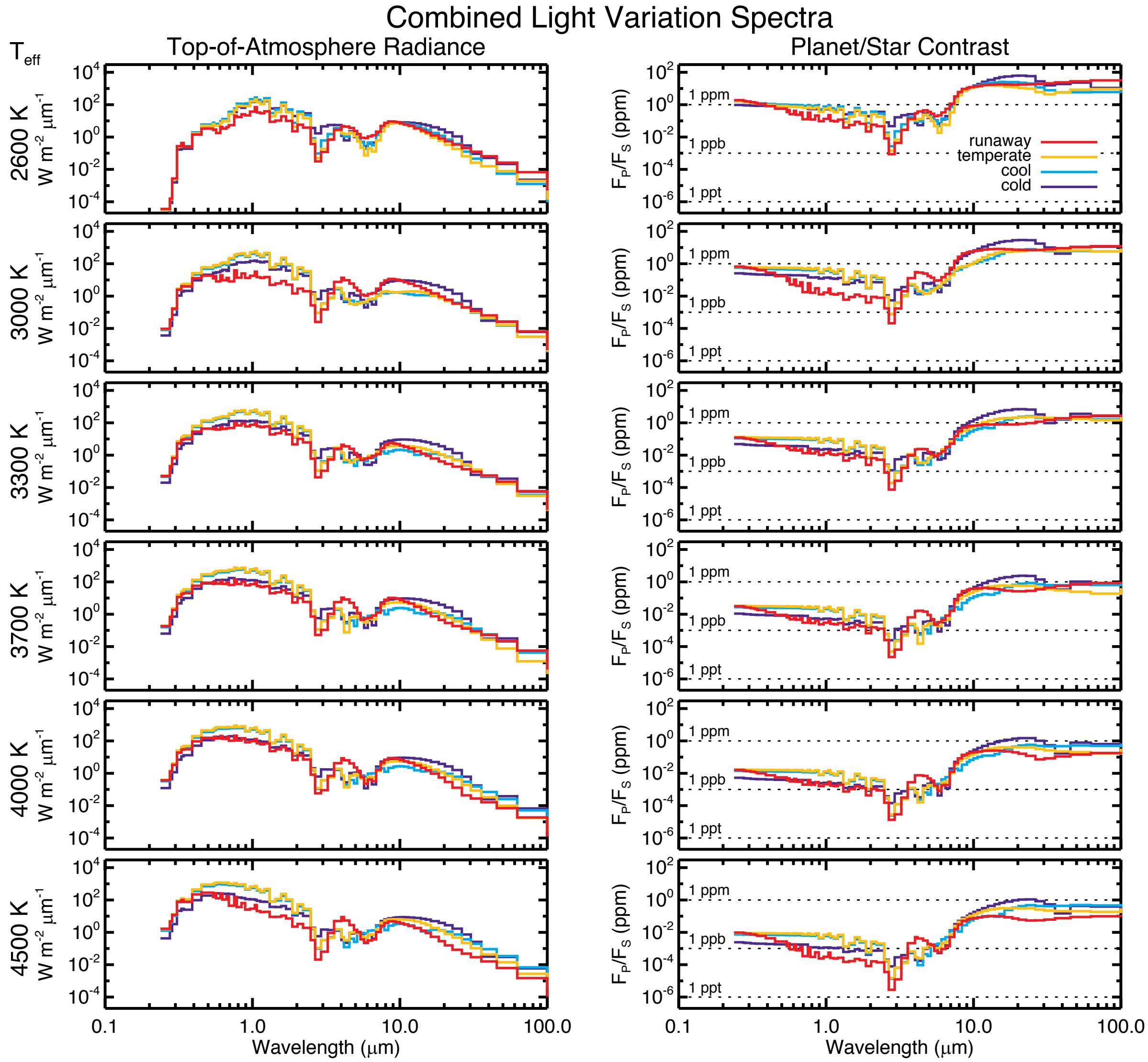